\documentclass[prx,reprint,twocolumn,showpacs,longbibliography]{revtex4-1}
\usepackage{amsmath,amssymb,bm,mathrsfs,graphicx, braket, times,amsthm,enumerate, scrextend}
\usepackage[colorlinks=true,citecolor=blue,linkcolor=blue]{hyperref}
\usepackage{longtable}
\usepackage{multirow}
\usepackage{array}
\usepackage{bigstrut}
\usepackage[all,cmtip]{xy}
\usepackage[normalem]{ulem}
\usepackage[usenames,dvipsnames]{color}
\usepackage{makecell}

\begin{document}
\title{Exhaustive List of Topological Hourglass Band Crossings in 230 Space Groups}

\author{Lin Wu}

\author{Feng Tang}
\email{Corresponding author (email): fengtang@nju.edu.cn}
\author{Xiangang Wan}
\affiliation{National Laboratory of Solid State Microstructures and School of Physics, Nanjing University, Nanjing, China}
\affiliation{Collaborative Innovation Center of Advanced Microstructures, Nanjing University, Nanjing, China}

\begin{abstract}
Topological semimetals with band crossings (BCs) near the Fermi level have attracted intense research activities in the past several years. Among various BCs, those enforced by an hourglass-like connectivity pattern, which are just located at the vertex in the neck of an hourglass and thus called hourglass BCs (HBCs),  show interesting topological properties and are intimately related with the space group symmetry. Through checking compatibility relations in the Brillouin zone (BZ),  we list all possible HBCs for all 230 space groups by identifying positions of HBCs as well as the compatibility relations related with the HBCs.
The HBCs can be coexisting with conventional topological BCs such as Dirac and Weyl fermions and based on our exhaustive list, the dimensionality and  degeneracy of the HBCs can be quickly identified. It is also found that the HBCs can be classified into two categories: one contains essential HBCs which are guaranteed to exist, while the HBCs in the other category may be tuned to disappear. Our results can help in efficiently predicting hourglass semimetals combined with first-principles calculations as well as studying transitions among various topological crystalline phases. 
\end{abstract}
\date{\today}
\maketitle

\section{Introduction}
¡¡¡¡In the past decade, there have been persistent and extensive research interest in  both theoretical and experimental studies on topological materials because of their fundamental and application value \cite{Qi-RMP, Hasan-RMP,Bansil-RMP,Ando-TCI,AV-RMP}. On one hand, theorists attempt to classify symmetry protected topological phases in a unified frame as far as possible
\cite{tenfold-1,tenfold-2,tenfold-3}. On the other hand, many kinds of topological materials, insulators or metals, were proposed with specific topological properties \cite{Qi-RMP, Hasan-RMP,Bansil-RMP,Ando-TCI,AV-RMP} and material realizations for them were oftentimes predicted through first-principles calculations and then verified by experiments \cite{Qi-RMP, Hasan-RMP,Bansil-RMP,Ando-TCI,AV-RMP}.

Among various topological materials, topological semimetals \cite{AV-RMP} with nontrivial band crossings (BCs) have attracted much attention since Weyl semimetals  were proposed and predicted to be able to host exotic surface states  and novel quantum responses \cite{Weyl-Wan,HgCrSe,TaAs-th-PRX,TaAs-th-NC,AV-RMP}. Other than Weyl semimetals where the two-fold Weyl BC can occur even without symmetry protection, various topological semimetals were proposed  where the relevant BCs are protected by crystallographic symmetry operations. With BCs distinguishable in dimensionality, degeneracy or connection pattern,  Dirac semimetals \cite{SMYoung,Na3Bi,Cd3As2,XYBi-DUYP},  topological semimetals with multifold-degenerate BCs \cite{Newfermions-Science,CoSi-th,RhSi-th,Weng-triple,DH-Nature}, nodal line semimetals \cite{Balents-NodalLine,RuiYu-NodalLine,Kane-NodalLine,ChenFang-NodalLine,CaTe-DUYP}, nodal-chain semimetals \cite{NodalChain}, Hopf-link semimetals \cite{Hopf-1,Hopf-2,Hopf-3,Hopf-4}  and so on, were born.

One type of topological semimetals, named by hourglass semimetals \cite{hourglass-Nature, Hourglass-Yaohong,ReO2,hourglass-YigeChen,hourglass-Ezawa,Hourglass-Takahashi,hex,tri,Hourglass-Wu} have attracted much attention recently. In an hourglass semimetal, the hourglass BC forms due to a hourglass-like band connectivity in the bulk Brillouin zone (BZ) as shown in Figs. \ref{5types}(a-e). To our best knowledge, material realizations of such hourglass fermions to date were mostly studied in orthorhombic crystal systems, for example Li$_3$(FeO$_3$)$_2$(SG 34, Pnn2) \cite{Li3(FeO3)2}, Ag$_2$BiO$_3$ (SG 52, Pnna ) \cite{Ag2BiO3-1, Ag2BiO3-2}, ReO$_2$ (SG 60, Pbcn)  \cite{ReO2}, SrIrO$_3$ \cite{hourglass-YigeChen}, Ta$_3$SiTe$_6$ and Nb$_3$SiTe$_6$ (SG 62, Pbmn) \cite{X3SiTe6}, AgF$_2$ (SG 61, Pbca) \cite{Dexi-AgF2,Tang-NP}, BaLaCuBO$_5$(SG 100, P4bm)\cite{BaLaCuBO5}.
It is worth pointing out that the hourglass BC can be topological BCs which have been extensively studied. For example,  the hourglass BC are allowed to own Weyl-like band topology, namely a finite monopole charge but such an hourglass Weyl point have to be located within special positions of the BZ. Other than hourglass Weyl point, the hourglass BC can also be a Dirac point. However, the hourglass  BCs needn't to be pinned down to high-symmetry points (HSPs) \cite{Newfermions-Science,SMYoung}, and can be movable in the high-symmetry line (HSL). Besides, the hourglass BCs could be robust against external perturbation as long as the perturbation preserves the symmetry, since the band inversion in the hourglass BC 
is usually enforced to occur  \cite{hourglass-Nature,Hourglass-Yaohong,hourglass-YigeChen,hourglass-Ezawa,ReO2}. The hourglass BC can also lie in  a nodal line. Furthermore, when each point of a nodal line is an hourglass BC, the nodal line is thus called hourglass nodal line. To date, the studies on hourglass semimetals are all based on specific kind of symmetry setting \cite{hourglass-Nature,Hourglass-Yaohong,hourglass-YigeChen,hourglass-Ezawa,ReO2}. However, an exhaustive study on hourglass BCs in all 230 space groups (SGs)  could  stimulate predicting and studying coexisting different types of topological hourglass BCs, and provide a complete searching or design principle for realistic hourglass materials as well.  

In this work, we use the compatibility relations in the 230 SGs as listed in the Bilbao server \cite{Bilbao} to obtain all possible and concrete positions in the BZ which allow hourglass BCs in the electronic band structures.   These results consider four settings (with/without  time-reversal symmetry (TRS) and with significant/negligible spin-orbit coupling (SOC)).   We find that  the hourglass dispersion can occur in all the crystal systems other than triclinic crystal system (since no HSL or high symmetry plane (HSPL) exists for triclinic crystal system). An interesting finding is that even in symmorphic SG, hourglass band structure also has a chance to appear.

In the main text the discussions  on hourglass BCs are  restricted to  the setting that TRS and significant  SOC are considered, while they are also applicable to the other three settings.  In Tables \ref{tab-1} and \ref{tab-2} of the main text, we list all the positions of essential hourglass nodal points and nodal lines in such setting, respectively. These essential results all belong to nonsymmorphic SGs.   It is worth pointing out that, with regard to essential hourglass BCs,  as along as a material crystallizes in SG as listed in Tables \ref{tab-1} and \ref{tab-2}, the hourglass band structure would always exist in the positions in the BZ shown in Tables \ref{tab-1} and \ref{tab-2}, regardless of the chemical composition. Hence, our results can be used to predict materials with topological hourglass excitations efficiently.

This paper is organized as follows. In Sec.~\ref{nece}, we first enumerate all possibilities or types of hourglass BCs based on which we propose  criteria or necessary conditions on compatibility relations or splitting pattern of existence for all types of hourglass BCs. Based on the necessary conditions, we obtain large number of possible hourglass BCs and list them in the Supplementary Material (SM) \cite{SM}. Then in Sec.~\ref{suff}, we further impose  an additional condition to obtain essential hourglass BCs. We list all  concrete positions in the BZ of essential hourglass nodal points and nodal lines  in Tables \ref{tab-1} and \ref{tab-2}, respectively.  Based on these tables, we  demonstrate three hourglass materials  whose electronic band structures are calculated by first-principles calculations in Sec.~\ref{material}. Conclusions and Perspectives are given in Sec.~\ref{conclusion}.

\section{Criteria of hourglass band crossing: necessary conditions}\label{nece}



\begin{figure*}[tbh!!]
	\begin{center}
		{\includegraphics[width=0.9 \textwidth]{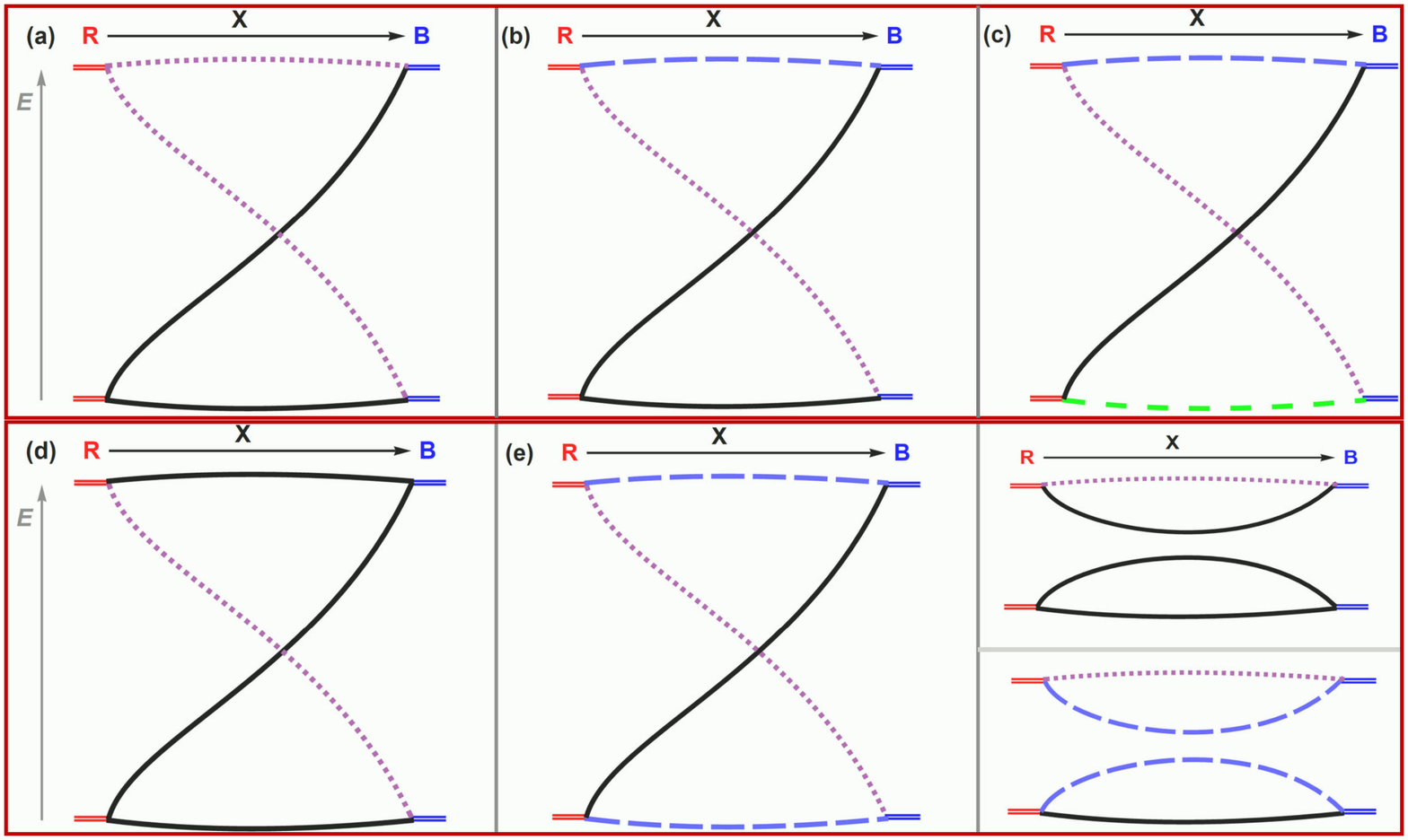}}
		\caption{Classification of the hourglass band structures: (a,b,c,d,e) correspond to hourglass BCs of type ($a$,$b$,$c$,$d$,$e$), respectively. R,B denote two (sets of) $\mathbf{k}$ vectors in the BZ which must be HSP or HSL while X, which can be HSL or HSPL, connect R and B. Energy bands from R to X split according to the compatibility relations. For stable band crossing as shown in the neck of hourglass, X must allow two different irreps as indicated by black solid and purple dotted lines. For types ($a$),($b$) and ($c$),  the numbers of additional involved irreps other than the two irreps in then neck for the bottom and top bands are 0,1,2, respectively, as indicated by different colors and styles of lines. It is easy to find that these hourglass band structures are stable against switching band orderings of R or B owing to different irreps in the bottom and top bands. However, the rest two hourglass structures in (d) and (e) can be tuned to gapped as shown in the last panel since the bottom and top bands share the same irrep. Note that for types-($a$,$b$,$c$),  once  for R and B, all possible splitting patterns in X are and can only be those as displayed, the hourglass band connectivity must exist since the partner switch of any Bloch states  have to happen to satisfy the compatibility relations,  thus they are essential.} \label{5types}
	\end{center}
	
\end{figure*}

In the BZ, any wave vector can be classified by its symmetry into an HSP, a point lying in an HSL or HSPL, and a generic point without any symmetry. Obviously, hourglass BC can only appear in HSL or HSPL for which two different irreducible representations (irreps) are allowed. Such HSL or HSPL hosting hourglass BC is denoted by X later. In the following we would outline all the necessary conditions for the formation of an hourglass BC.

 First of all, we describe the property of wave vectors related with any hourglass BC. Any hourglass structure consists of four sets of energy bands in X as shown in Figs. \ref{5types}(a-e): the four sets of bands connect two energy levels at R and B in a zigzag manner. Here R or B are connected by X, providing two energy levels splitting in X. Thus for each hourglass BC, R-X-B should be specified.  All possible combinations of R-X-B are given in Table \ref{RXB}. Many examples of R-X-B can be found in Secs. II-V of the SM \cite{SM}: for example, with respect to T-K-GM in SG 32, where T=$(0,\frac{1}{2},\frac{1}{2})$, K=$(0,v,w)$ and GM=(0,0,0), it is easy to find that K connects T and GM. Note that we exactly follow the name convention of wave vectors in the Bilbao server \cite{Bilbao} where all the coordinates in the conventional basis can also be found. Interestingly, when X is an HSPL, there would be an infinite number of pathes in X connecting R and B, each of which contributes an hourglass BC, thus resulting in an hourglass nodal line. However, when X is a HSL, we should pay some attention:

1. If no HSPL contains X,  the hourglass BC in X must be an (symmetry-allowed) hourglass nodal point, that is to say, the band touching in hourglass BC in X cannot survive in any direction away from the BC;

2. On the contrary, if we can find one or several HSPLs containing  X, then we should check  that if there exist(s) one or several HSPLs containing X satisfying that, the two different irreps of the hourglass BC  can still maintain to be two different irreps in these HSPLs.  If that is true, the hourglass BC in X would lie in a nodal line in corresponding HSPLs. Otherwise, the hourglass BC in X is an hourglass nodal point.

Thus for X being an HSL, it is still possible to obtain a nodal line threading the hourglass BC in X. We find that such case usually (but not always) indicates existence of another HSPL, denoted by X$'$, hosting a hourglass nodal line and X $\in$ X$'$. For example, from Sec. II of the SM \cite{SM}, SG 62 could host an hourglass BC in D (an HSL) connecting S and X. However, we further find that such a BC actually lies in a nodal line within an HSPL, L (and D $\in$ L). In fact, L, itself can host an hourglass nodal line.
However, such X$'$ not always exists, with the counterexample being  SG 205 \cite{SM}:  the only possible hourglass BC in an HSL, ZA $\in$ B (an HSPL), is found to lie in a nodal line within B, but B cannot host hourglass nodal line.


An important characteristic of  hourglass structure is that the band inversion of a BC arises from a state-switching process from R to B through X. This can be illustrated in detail as follows. Assume that there has existed an hourglass BC in R-X-B, so that the properties it has thus provide necessary conditions for realize hourglass BCs. Since the little group of X must be a subgroup of R or B, the energy level at R or B may split in X, according to the compatibility relations of irreps between R or B and X \cite{Bilbao}.  It is easy to find that R and B should allow degenerate energy level which is able to split to two and only two sets of bands, corresponding to two irreps in X (note that these two irreps may be the same one).   We emphasize that we needn't care about which irrep at R and B of the aforementioned energy level splits to irreps in X and what matters essentially is that how bands originated from R or B split in $X$, called splitting pattern, which can be obtained from the compatibility relation given in the Bilbao server \cite{Bilbao}. As shown Figs. \ref{5types}(a-e), two energy levels at R and B split in X,  we can thus use the notation as:\\
\begin{equation}\label{splitting}
  \mathrm{X}_i\oplus\mathrm{X}_j,\mathrm{X}_{i'}\oplus\mathrm{X}_{j'};\mathrm{X}_{i''}\oplus\mathrm{X}_{j''},\mathrm{X}_{i'''}\oplus\mathrm{X}_{j'''},
\end{equation}
to specify the splitting pattern in any hourglass structure where  $i,j,i',j',\ldots$  denote the irreps in X. The two splitting patterns before (after) the semicolon correspond to higher and lower energy levels  at R (B), respectively. To be specific,
  $\mathrm{X}_i\oplus\mathrm{X}_j$ means that the higher energy level at R split to irreps X$_i$ and X$_j$, $\mathrm{X}_{i'}\oplus\mathrm{X}_{j'}$ means that the lower energy level at R split to irreps X$_{i'}$ and X$_{j'}$,  $\mathrm{X}_{i''}\oplus\mathrm{X}_{j''}$ means that the higher energy level at B split to irreps X$_{i''}$, and X$_{j''}$ and $\mathrm{X}_{i'''}\oplus\mathrm{X}_{j'''}$ means that the lower energy level at B split to irreps X$_{i'''}$ and X$_{j'''}$.\\

Note that the irreps in X participating in an hourglass structure cannot be chosen arbitrarily and should satisfy certain constraints as shown below. First of all, $\mathrm{X}_i\oplus\mathrm{X}_j$ cannot be equal to $\mathrm{X}_{i''}\oplus\mathrm{X}_{j''}$ otherwise they could be connected with each other violating the hourglass band connectivity. So  $\mathrm{X}_i\oplus\mathrm{X}_j$ $\ne$ $\mathrm{X}_{i''}\oplus\mathrm{X}_{j''}$, thus the states from lower energy should have to be incorporated, as shown later, to result in a switch of different irreps in X, which enforces an hourglass BC.

\begin{table}
  \centering
  \caption{All possible combinations of R, X and B which can host hourglass band crossings in X connecting R and B. The nodal line among the table may not be the hourglass structure because it is possible that not all the points constituting the nodal line are originated from  hourglass crossings.  }\label{RXB}
\begin{tabular}{c|c|c|c}
  \hline\hline
  R & X & B & Property of HBC\\\hline
  HSP & HSL & HSP & \makecell{Hourglass nodal point, or \\ Lying in a nodal line} \\\hline
   HSP & HSPL & HSP& Hourglass nodal line \\\hline
  HSP & HSPL & HSL& Hourglass nodal line \\\hline
  HSL & HSPL & HSL&  Hourglass nodal line\\
  \hline\hline
\end{tabular}
\end{table}


Next, we list all possible cases of the splitting patterns in X for an hourglass BC in R-X-B shown in Figs. \ref{5types}(a-e). With no loss of generality, we label the different irreps for the BC in the neck of the hourglass as X$_1$ and X$_2$, which are represented by black solid line and purple dashed line, respectively, as in Figs. \ref{5types}(a-e). Then considering the irreps for the bottom and top part of the hourglass, we obtain all the five possibilities as follows:\\

{\bf Type {\emph{a}}:}\qquad The irreps  of the bottom and top parts are different but they share the same irreps as those of the neck parts. We can denote the top irrep as X$_2$ as in Fig. \ref{5types}(a). Note that the other choice of the top irrep being X$_1$ would give the same type of hourglass (just reversing the order of R and B).  Hence, the splitting patterns can be denoted as: 2X$_2$$,$2X$_1$;X$_1$$\oplus$X$_2$, X$_1$$\oplus$X$_2$. It is easy to find that X$_1$ and X$_2$ have been interchanged between R and B by this formal notation: The higher energy levels of R and B can not be directly connected since they contain different irreps in X (i.e. 2X$_2$$\neq$X$_1$$\oplus$X$_2$) as well as the lower energy levels. The states in  the higher and lower energy level at R interchange X$_1$ and X$_2$ in X and finally change to those at B.

{\bf Type {\emph{b}}:}\qquad The same as type $a$, the irreps of the bottom and top parts are different but they only share one common irrep with those of the neck parts. Denote the common irrep as X$_1$ for the bottom band, and the other irrep for the top band is X$_3$. The splitting patterns are thus
X$_2$$\oplus$X$_3$,2X$_1$;X$_1$$\oplus$X$_3$, X$_1$$\oplus$X$_2$. From this formal notation, X$_1$ and X$_2$ is interchanged from R and B through X, which are just the irreps participating  in the hourglass BC.

{\bf Type {\emph{c}}:}\qquad The irreps of the bottom and top parts are different and they only share no common irrep with those of the neck parts. Denote the irrep as X$_4$ for the bottom band, and the other irrep for the top band is X$_3$. The splitting patterns are thus
X$_2$$\oplus$X$_3$,X$_1$$\oplus$X$_4$; X$_1$$\oplus$X$_3$, X$_2$$\oplus$X$_4$.

{\bf Type {\emph{d}}:}\qquad Different from the above three types, for this type, the bottom and top bands share the same irrep in X which is X$_1$ or X$_2$. We can denote the irrep of the bottom and top bands as X$_1$.  Thus the splitting patterns are X$_1$$\oplus$X$_2$,2X$_1$;2X$_1$, X$_1$$\oplus$X$_2$.

{\bf Type {\emph{e}}:}\qquad  The bottom and top bands share the same irrep in X which is not X$_1$ or X$_2$. We can denote the irrep of the bottom and top bands as X$_3$.  Thus the splitting patterns are X$_2$$\oplus$X$_3$,X$_1\oplus$X$_3$ and  X$_1$$\oplus$ X$_3$, X$_2$$\oplus$X$_3$.

The hourglass BCs of the above five types are schematically shown in Figs. \ref{5types}(a,b,c,d,e), respectively. The degeneracy of the hourglass BC is simply the sum of dimension of the two interchanged irreps X$_1$ and X$_2$. Different from the hourglass BCs of types $a$-$c$, the hourglass BCs of types $d$ and $e$ can be gapped by interchanging the band ordering in R or B. For example, with respect to hourglass BC of type $d$, when interchanging the band ordering at R, the splitting patterns are thus:  2X$_1$, X$_1$$\oplus$X$_2$;2X$_1$, X$_1$$\oplus$X$_2$. Hence the higher and lower energy levels can both be directly connected from R and B through X  without any state switch and a continuous energy gap could exist to separate the four sets of bands. The effect of interchanging energy levels R or B or both R and B for all the 5 types giving  equivalent hourglass band structures, are displayed in Fig. 1 of the SM \cite{SM}.

\begin{figure}
  \includegraphics[width=0.45 \textwidth]{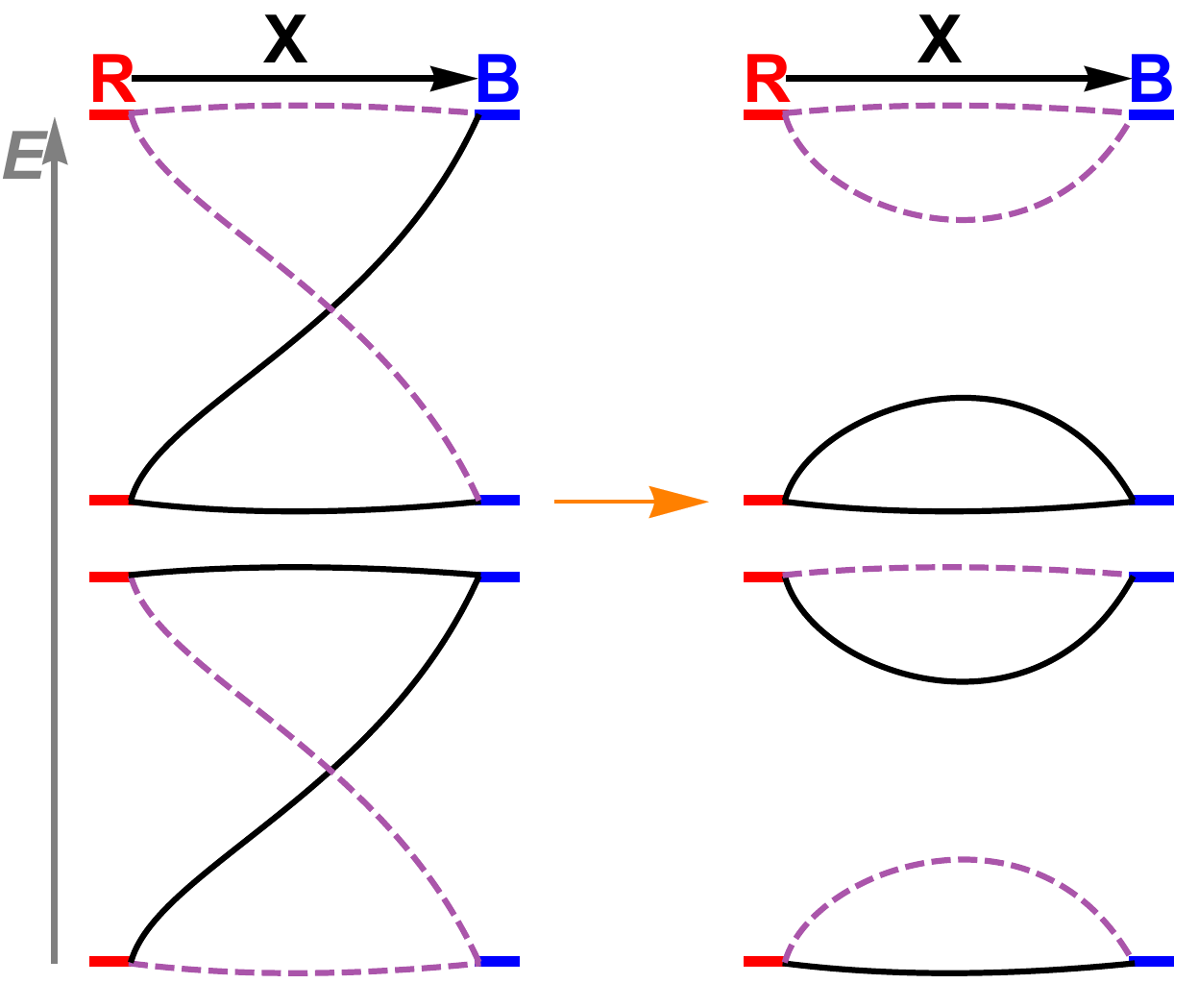}\\
  \caption{When two hourglass structures both of type $a$ (see Fig. \ref{5types}) have existed, they can still be turn to gapped as shown in the right panel through tuning the band orderings. }\label{gap}
\end{figure}

It is also worth pointing out that even for types ($a,b,c$)  shown in Fig. \ref{5types}, the hourglass BCs can be tuned to disappear when other kind of splitting pattern(s) with common irrep(s) shared in the hourglass structure are incorporated. Here we show a specific example of this situation shown in Fig. \ref{gap}  where two hourglasses in type $a$ are formed in the left panel but they can be tuned to gapped as shown in the right panel. As a matter of fact, it is found that although the lower hourglass satisfies compatibility relations or splitting patterns for hourglass BC of type $a$, R can allow another splitting pattern which could accommodate the bands originated from B without any band crossings. This is understood that what we proposed are necessary conditions given the hourglass BC has been formed, while they can also be gapped or coexist with other kinds of band connectivity.

Based on the necessary conditions for types $a$-$e$, we exhaustively investigate the compatibility relations listed in  the Bilbao server \cite{Bilbao} and obtain comprehensive results all listed in Secs. II,III,IV and V of the SM \cite{SM}, including the SG, R-X-B, related splitting pattern as Eq. \ref{splitting}, dimensionality (nodal line or point) and the type. It is interesting to find that there are no results for type $b$ and type $d$ hourglass BCs in the symmetry setting adopted in the main text. Even though those results may not essentially (i.e. accidently ) guarantee an hourglass structure, they can indeed provide possibility of an hourglass BC and can also aid in designing structures or applying external perturbations to tune hourglass structure(s) to be gapped or even transform to other types of BCs.

\section{Sufficient condition for essential hourglass band crossing and results}\label{suff}

The tables of Secs. II,III,IV and V of the SM \cite{SM} contain many essential hourglass BCs (as printed in red there), which can be applied to predicting materials with topological hourglass excitations near the Fermi level.  In this section, we describe our  additional condition other than the above necessary conditions, which could guarantee that the hourglass BC is essential, that is, they must exist, cannot be gapped and are the only kind of possible band connectivity. To obtain an essential hourglass BC in R-X-B, obviously, it should allow splitting patterns in type $a$, $b$ or $c$ since hourglass BCs of types $d$ and $e$ can be tuned to be gapped as shown in the last panel of Fig. \ref{5types}.
Furthermore the following condition should be satisfied:\\

\textit{No other splitting patterns exist other than those required for type $a$, $b$ or $c$.}\\

By the above additional requirement, we filter out  nonessential hourglass BCs and those essential hourglass nodal points and nodal lines and they are collected in Tables \ref{tab-1} and \ref{tab-2}, respectively, where only X is specified.  The concrete coordinates of X as well as the splitting patterns (so that the degeneracy of hourglass BC can be obtained) can be found in the SM \cite{SM}.

Here we take SG 182 as an example to illustrate the essential hourglass BC. By check all the compatibility relations for SG 182 listed in the Bilbao sever \cite{Bilbao},
we can see that there is an essential hourglass nodal point of type $a$ on the high-symmetry line U which connects two high-symmetry points L and M. Here we label these irreps with the same convention in Ref. \cite{Bilbao}. The irreps of U include two one-dimensional irreps U$3$ and $\mathrm{U}4$. Meanwhile, there are two two-dimensional irreps $\mathrm{L}{2,4}$  (here 2,4 means that the irreps 2 and 4 for L are paired by TRS to give the coirrep) and $\mathrm{L}{3,5}$ at point L and there is only one two-dimensional irrep $\mathrm{M}5$ at point M. The  irreps at L are found to be related with those in U according to the compatibility relations as follows \cite{Bilbao}:

\begin{equation}\label{eq1}\begin{split}
\mathrm{L}{2,4}\rightarrow 2\mathrm{U}3,\\
\mathrm{L}{3,5}\rightarrow 2\mathrm{U}4,\\
\end{split}\end{equation}
while for M, we have \cite{Bilbao},
\begin{equation}\label{eq2}
\mathrm{M}5\rightarrow \mathrm{U}3\oplus\mathrm{U}4.
\end{equation}
 The above compatibility relations obviously satisfy the requirement of hourglass BC of type $a$.  Furthermore, since all possible compatibility relations  for L and U as well as M and U are shown above, thus there is only one (inequivalent) splitting pattern for L-U-M. Therefore such hourglass BC is essential and furthermore found to be an hourglass nodal point. We thus list U in Table \ref{tab-1} and assign 182 to it meaning that U in SG 182 can host essential hourglass nodal point. It can also be found that other than 182, several other SGs such as SGs 4, 90, 94 could also host essential hourglass nodal point in U as read from Table \ref{tab-1}. In Sec. \ref{sectioniva}, the electronic band structure of a realistic material in SG 182 is calculated by first principles where the essential hourglass nodal  points are explicitly demonstrated.

As shown in Tables \ref{tab-1} and \ref{tab-2}, a notable feature of the essential results is that all the nodal points except in LD and V and all the nodal lines are of type $a$, i.e. only two different irreps are allowed in X. This is understandable since many HSLs or HSPLs can only host two different irreps at most.  We also note that 12 SGs can host essential hourglass nodal points in more than one HSLs or HSPLs and 19 SGs host essential nodal lines in more than one HSPLS, for example, SG 118 can host essential hourglass nodal lines in HSPLs B and F, so that the positions of those multiple nodal lines or loops in the BZ may affect the topological structure sensitively. Besides, there are 23 chiral SGs which are found to be able to host essential hourglass nodal points, for example, SG 90, could host a finite monopole charge \cite{KramersWeyl}.

\begin{table}
  \centering
   \caption{All positions of HSLs for essential hourglass nodal points. The first column contain labels for the HSLs (X) and in the second column, it contains several SG numbers which could allow essential hourglass nodal points in the corresponding HSLs. The letter a and c denote the essential hourglass nodal points with type $a$ and type $c$ BC, respectively. Chiral SGs are printed in bold style and the SGs are in red to which the materials discussed in the main text belong. }\label{tab-1}
  \begin{tabular}{c|c}
    \hline
    X& SG\\\hline \hline
    LD & $a$: \textbf{4},\textbf{17},\textbf{19},\textbf{20}; $c$: \textbf{77},\textbf{80},\textbf{93}, \textbf{94},\textbf{98} \\\hline
    U & $a$:\textbf{4}, \textbf{90}, \textbf{94}, 113, 130, 138, \textbf{169}, \textbf{170}, \textbf{173}, \textbf{178}, \textbf{179},{\color{red} \textbf{182}}\\\hline
     V & $a$: \textbf{4}, 106, 133, 219;$c$:\textbf{77},\textbf{93}\\\hline
      W & $a$:\textbf{4}, \textbf{76}, \textbf{78}, \textbf{91}, \textbf{95}, 120\\\hline
       G & $a$:\textbf{17}\\\hline
        H & $a$:\textbf{17},\textbf{20},60\\\hline
         Q & $a$:\textbf{17}\\\hline
          A & $a$:\textbf{18},54,56\\\hline
           B &$a$:\textbf{18},56\\\hline
            DT &$a$:\textbf{18}, \textbf{19}, \textbf{90}, \textbf{92}, \textbf{94}, \textbf{96}, 113, 114\\\hline
             SM &$a$:\textbf{18}, \textbf{19}\\\hline
              D &$a$:\textbf{20},{\color{red}36},{\color{red}52}\\\hline
               P &$a$:45\\\hline
               E &$a$:54\\\hline

    \hline
  \end{tabular}

\end{table}

\begin{table}
  \centering
    \caption{All positions of HSPLs for essential hourglass nodal lines. The first column contain labels for the HSPLs (X) and in the second column, it contains several SG numbers which could allow essential hourglass nodal lines in the corresponding HSPLs. The  letter a means that the essential hourglass nodal line is within type $a$ hourglass BC.  SG 62 (printed in red) is used to show an hourglass nodal line material in the main text. }\label{tab-2}
  \begin{tabular}{c|c}
    \hline
    X& SG\\\hline \hline
    F & $a$: 7, 102, 104, 118 \\\hline
    G & $a$:7\\\hline
     B & $a$: 9, 45, 46, 100, 102, 104, 106, 108, 110, 117, 118, 120\\\hline
      K & $a$:26, 30, 31, 32, 33, 34, 36\\\hline
       L & $a$:26, 29, 30, 33, 34, 60, 61, {\color{red}62}\\\hline
        M & $a$:28, 29, 31, 32, 33, 34, 40, 41\\\hline
         N & $a$:28, 29, 30, 31, 34, 57, 61\\\hline
          P & $a$:39, 41\\\hline
           Q &$a$:39, 41\\\hline
            E &$a$:43\\\hline
             J &$a$:43\\\hline
              A &$a$:45, 109, 110, 122\\\hline
               W &$a$:60,61\\\hline
               D &$a$:158,186,188\\\hline
                C& $a$:159, 161, 185, 190, 220          \\
    \hline \hline
  \end{tabular}

\end{table}
Besides, as shown in Tables \ref{tab-1} and \ref{tab-2}, essential hourglass nodal line can occur in the crystal systems other than triclinic crystal system, and while essential hourglass nodal point can emerge in  crystal systems other than triclinic and trigonal systems. This indicates that hourglass materials could be ubiquitous in nature.

\section{Material Representatives}\label{material}

\subsection{Materials with essential hourglass nodal points}\label{sectioniva}
\begin{figure*}[tbh!!]
	\includegraphics[width=1 \textwidth]{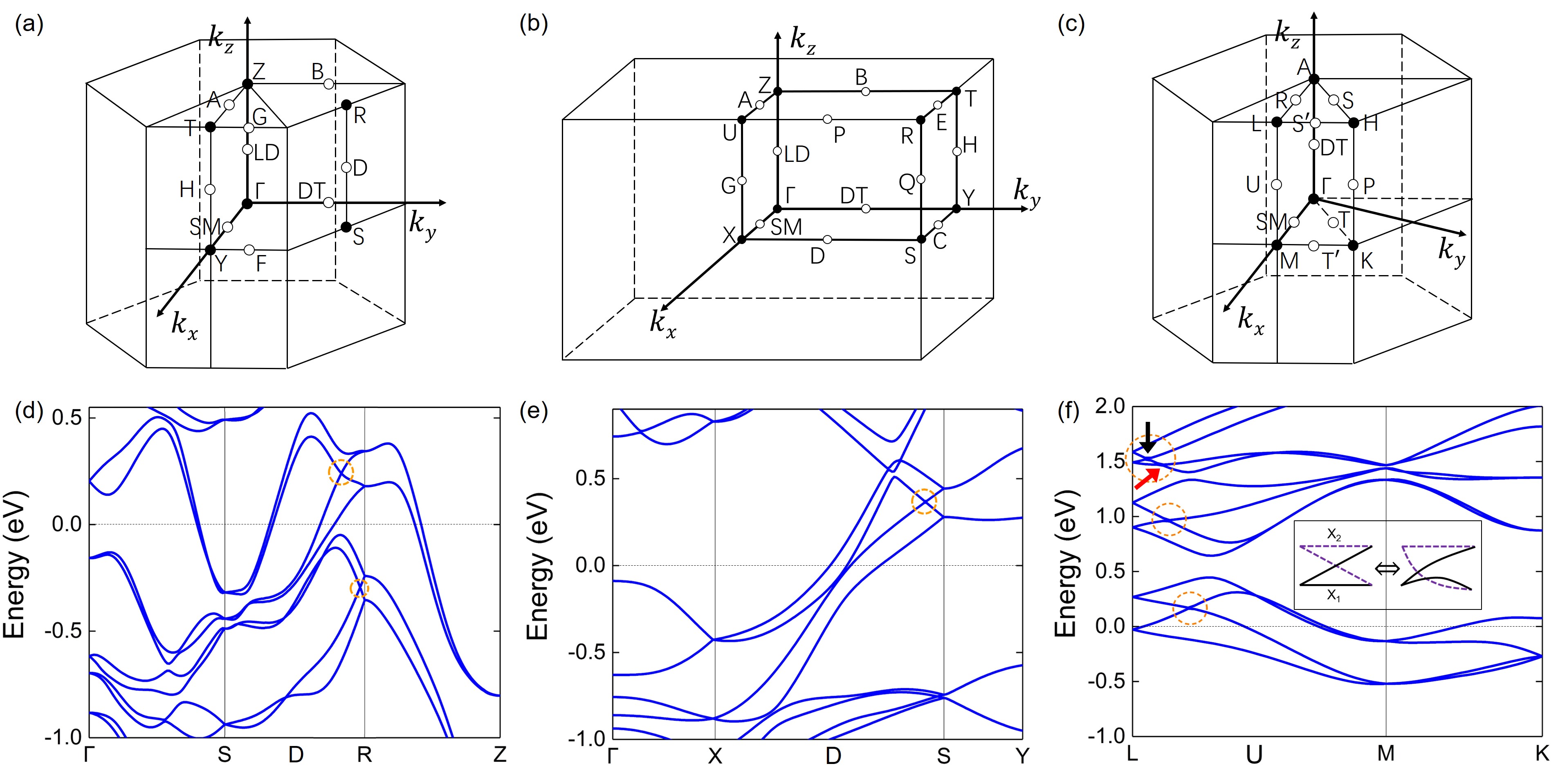}\\
	\caption{(a) The BZ of base-centred orthorhombic lattice, where the labels of inequivalent HSPs and HSLs are given, following the notation adopted in the Bilbao server \cite{Bilbao}. (b) The BZ of primitive orthorhombic lattice, where the labels of inequivalent HSPs and HSLs are given, following the notation adopted in the Bilbao server \cite{Bilbao}. (c)The BZ of the trigonal lattice where we label all the inequivalent HSPs and HSLs following the Bilbao Server \cite{Bilbao}. (d) Electronic band structure of AsPb$_2$Pd$_3$ (SG 36) by first principles calculations where S-D-R line could always host hourglass nodal points. S and R are two HSPs and D is HSL. Two such nodal points are circled by dashed curves almost near the Fermi level. (e)The first-principles calculated electronic band structure  of Sr$_2$Bi$_3$(SG 52) with the hourglass nodal point indicated by orange dashed circle . (f) The principle-calculated electronic band structure of ReO$_3$ (SG 182) with three Weyl hourglass points indicated by dashed circles. Note that, for the highest circle, there exist two BCs, one (denoted by black arrow) is caused by hourglass structure and robust while the other one (denoted by the red arrow) is not robust and can be tuned to disappear as shown in the inset.}\label{fig-mat1}
\end{figure*}

As shown in Table \ref{tab-1}, for SG 36, D connecting HSPs S and R as shown in Fig. \ref{fig-mat1}(a), could host essential hourglass nodal point. The irreps of D are two one-dimensional irreps $\mathrm{D}{3}$ and $\mathrm{D}{4}$.  Irreps at R or S are related with those in D as indicated by the compatibility relations.  There are two two-dimensional irreps at point R which are reduced  to $2\mathrm{D}{3}$ and $2\mathrm{D}{4}$.  As R point, S point also allows two two-dimensional irreps, but they are both reduced  to $\mathrm{D}{3} \oplus \mathrm{D}{4}$. Thus these splitting pattern satisfies the requirement of type $a$ hourglass BC. Furthermore, the hourglass BC in D is essential and found to be just an hourglass nodal point \cite{SM}.

Similarly for SG 52,  D listed in Table \ref{tab-1}, connects HSPs S and X as shown in Fig. \ref{fig-mat1}(b), could also host essential hourglass nodal point as analyzed below. The irreps of D are two two-dimensional irreps $\mathrm{D}{2,4}$ and $\mathrm{D}{3,5}$. Two different splitting patterns for S to X are $2\mathrm{D}{2,4}$ and $2\mathrm{D}{3,5}$ while the only splitting pattern for point X to D is $\mathrm{D}{2,4} \oplus \mathrm{D}{3,5}$. Therefore hourglass BC of type $a$ could occur in D and the BC is found to be essential as in SG 36.

The above two SGs both own achiral operations,  while  SG 182 which is chiral, could host chiral hourglass nodal point in HSL U as shown in Table \ref{tab-1}: from L to U (see Fig. \ref{fig-mat1}(c)), there are two possible splitting patterns $2\mathrm{U}{3}$ and $2\mathrm{U}{4}$, while from M to U (see Fig. \ref{fig-mat1}(c)), the only splitting pattern is $\mathrm{U}{3} \oplus \mathrm{U}{4}$, as shown in Eqs. \ref{eq1} and \ref{eq2}. Hence, an hourglass BC of type $a$ could occur in U and is further found to be essential. As SG 182 is a chiral SG and U3 and U4 are both one-dimensional, then such hourglass nodal point in U is a Weyl point.\\

First we showcase AsPb$_2$Pd$_3$\cite{AsPb2Pd3} crystallizing in SG 36 with base-centred orthorhombic lattice, whose first-principles calculated electronic band structure are shown in Fig. \ref{fig-mat1}(d). As discussed above, the bands in S-D-R must host an hourglass nodal point. Clearly, the electronic band  structure in Fig. \ref{fig-mat1}(d) demonstrates two such hourglass nodal points as indicated by dashed circles, which are very near to the Fermi level. We should point out that  although SG 36 is not centrosymmetric, the two-fold degenerate nodal points in D are not Weyl points since D owns mirror symmetry m$_{010}$ which guarantees vanishing Chern number around the nodal points. Then we present Sr$_2$Bi$_3$\cite{Sr2Bi3} as an hourglass  material example, which crystallize in SG 52 owning inversion center with primitive orthorhombic lattice. The BZ is shown in Fig. \ref{fig-mat1}(b), and the electronic band structure by first-principles calculations is shown in Fig. \ref{fig-mat1}(e).  Because SG 52 owns an inversion center, each band should be double-fold degenerate.   The four-fold degenerate essential hourglass nodal point in D is indicated  by an orange dashed circle.

Other than achial hourglass materials as above, for SG 182 we present ReO$_3$ \cite{ReO3}  which could host chiral and essential hourglass nodal point. Such hourglass nodal point must appear in HSL U connecting L and M  shown in the BZ of Fig. \ref{fig-mat1}(c). The electronic band structure of ReO$_3$ by first principles calculations is displayed in Fig. \ref{fig-mat1}(f) where we highlight three such hourglass Weyl points by dashed circles. Note that for the highest dashed circle, there exists an additional BC (denoted by an red arrow) other than the hourglass one  (denoted by the black arrow) in the hourglass structure in the band plot. Such a BC is not robust to external strain unlike the hourglass BC and can be tuned to be gapped easily  as schematically shown in the inset of Fig. \ref{fig-mat1}(f).

\subsection{Material with essential hourglass nodal line}
We have listed all positions of essential hourglass nodal lines in Table \ref{tab-2}. As shown in Table \ref{tab-2}, for SG 62 with primitive orthorhombic lattice there is an hourglass nodal line in  L($k_x=\frac{\pi}{a}$) plane (i.e. UXSR plane in Fig. \ref{fig-mat1}(b)). As an example we show that Ba$_2$ReO$_5$ in SG 62 could host such hourglass nodal line. We find that S-D-X and S-Q-R could host essential hourglass BCs where D and Q are both HSLs. However, we find that the hourglass BC in D or Q actually lies in a nodal line in L plane. Actually, S-L-G, S-L-P, S-L-R,S-L-U,S-L-X are all found to display essential hourglass nodal lines by our criteria described in Secs. \ref{nece} and \ref{suff}. Above all, we expect a nodal loop around S in plane L to emerge.
\begin{figure*}[tbh!!]
	\centering
	\includegraphics[width=1 \textwidth]{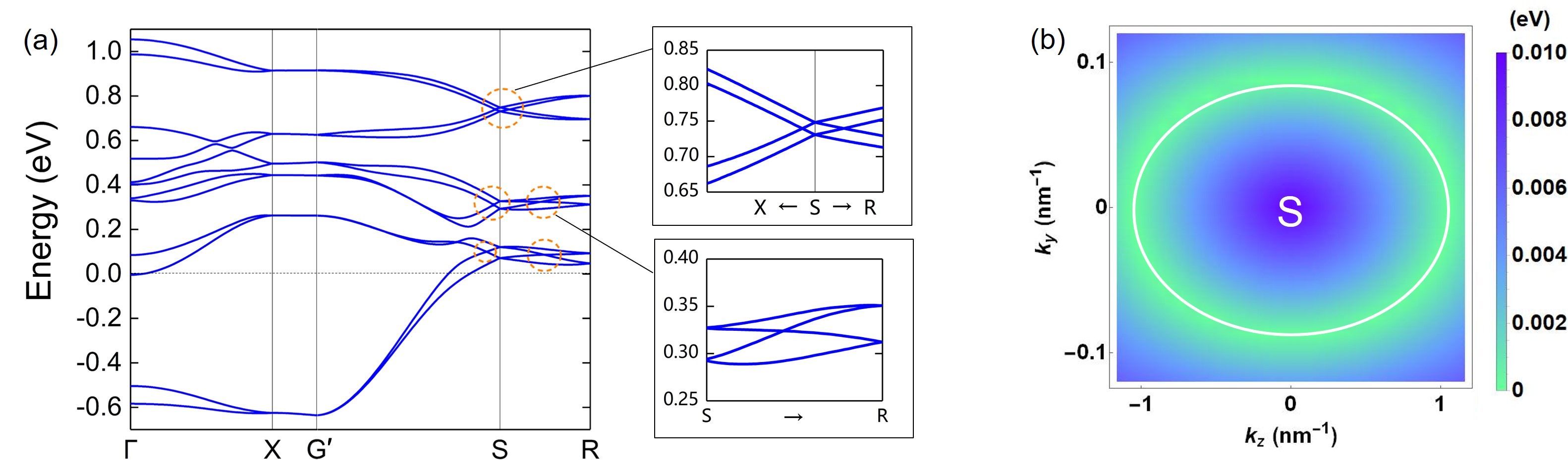}\\
	\caption{(a) The electronic band structure of Ba$_2$ReO$_5$ where G$'$ is the middle point of UX. Note that the bands in X-G$'$ are always four-fold degenerate since only one 4 dimensional irrep is allowed in HSL G. In fact, any path (even not line) connecting S and any point in HSL G  could host essential hourglass BC. We highlight six hourglass BCs in the band structure with two of them are enlarged shown in the right-side insets; (b) The expected hourglass nodal loop (the white loop) around S in L plane is verified by our first principles calculations. }
	\label{fig-mat2}
\end{figure*}
The first-principles calculated band structure of Ba$_2$ReO$_5$\cite{Ba2ReO5} is shown in Fig. \ref{fig-mat2}(a) where six hourglass BCs near the Fermi level are shown.  Any energy band is double-fold degenerate for SG 62 is centrosymmetric.  Besides,  the bands in HSL G (G$'$X)  are always four-fold degenerate \cite{Bilbao} as shown in the first principles calculated band structure in Fig. \ref{fig-mat2}(a). The path connecting S and any point in G could host an hourglass BC, which finally could  constitute a Dirac nodal line. As displayed in Fig. \ref{fig-mat2}(b), such a Dirac nodal loop obtained through first-principle calculations is around S in the L plane. Furthermore, it is found that the Dirac nodal loop own a very narrow band width being only about 13.6 meV based on our first-principles results. It is also worth mentioning that  any path in L which connects S and arbitrary point in P can host an hourglass BC, which is shown in the SM \cite{SM}.

\section{Conclusions and Perspectives}\label{conclusion}
Based on the compatibility relations in the Bilbao server \cite{Bilbao} we obtain all the possible hourglass band crossings in the BZ of all 230 SGs. We consider not only double valued representations but also single-valued representations. They are suitable for electronic materials with  significant spin-orbit coupling, and  negligible spin-orbit coupling or bosonic systems (e.g. phonon, photon or magnon systems), respectively. Furthermore, we also consider situations with/without time-reversal symmetry for nonmagnetic/magnetic materials.  We  highlight hourglass band crossings which are essential for they are very promising to be used to realize materials with hourglass band crossing near the Fermi level. 
The space groups where the essential hourglass structure exists are all nonsymmorphic groups. Based on the essential results, we predict several hourglass materials which are expected be synthesized for further studies.

It is worth pointing out that several other special occasions (for example, R and B are related by symmetry, or even different by a reciprocal lattice vector) which can also host hourglass band crossings are not captured in our search  since their compatibility relations  are not explicitly listed in the Bilbao server \cite{Bilbao}. However, our methodology can be directly applied to these cases. Although we  apply our results to show hourglass band structure in nonmagnetic materials, but they can also be applied to magnetic materials, or even bosonic systems. Besides, since layer group or rod group is a subgroup of space group, our results can also be used to search hourglass materials in lower dimension.

\section{Acknowledgements}
We were supported by the National Key R$\&$D Program of China (grants 2017YFA0303203 and 2018YFA0305704), the National Natural Science Foundation of China (NSFC; grants 11525417, 11834006, 51721001 and 11790311) and the Excellent Programme at Nanjing University. XW also acknowledges the support from the Tencent Foundation through the XPLORER PRIZE. FT thanks Prof. Dingyu Xing and Prof. Baigen Wang for their kindly and substantial support on scientific research.
\bibliography{biblio0117}

\clearpage

\onecolumngrid

{\centering{\bf Supplementary Material for ``Exhaustive List of Topological Hourglass Band Crossings in 230 Space Groups''}}

\author{Lin Wu}
\author{Feng Tang}\email{Corresponding author (email): fengtang@nju.edu.cn}
\author{Xiangang Wan}
\affiliation{National Laboratory of Solid State Microstructures and School of Physics,
Nanjing University, Nanjing 210093, China}
\affiliation{Collaborative Innovation Center of Advanced Microstructures, Nanjing
University, Nanjing 210093, China}

\date{\today}

\maketitle
\setcounter{section}{0}
\setcounter{figure}{0}
\setcounter{equation}{0}
\section{Preliminary remarks}

The purpose of our work is to exhaustively search for all possible hourglass band crossings in 230 space groups. Using all the compatibility relations listed in the Bilbao server\cite{Bilbao} and proposed necessary conditions  in Sec. II of the main text  (where all the hourglass band crossings are classified  into five types, namely types-$a$, $b$, $c$, $d$ and $e$),  we obtain all possible positions of hourglass band crossings as well as the concrete relevant splitting patterns.   In Tables II and III of the main text, we  list all the positions of essential hourglass nodal points and nodal lines, considering time-reversal symmetry and significant spin-orbit coupling. In this Supplementary Material, we display all the positions of hourglass band crossings for  all the four settings that with/without TRS and with neglected/significant SOC in detail by four sections (corresponding to Secs. \ref{trssoc},\ref{trsnosoc},\ref{notrssoc},\ref{notrsnosoc} of the Supplementary Material, respectively):

\begin{itemize}
  \item In Sec. \ref{trssoc}, we give the results for time-reversal symmetric (TRS) electronic systems with significant spin-orbit coupling (SOC);
  \item In Sec. \ref{trsnosoc}, we give the results for TRS electronic systems with neglected SOC;
  \item In Sec. \ref{notrssoc}, we give the results for TR-broken (TRB) electronic systems with significant SOC;
  \item In Sec. \ref{notrsnosoc}, we give the results for TRB electronic systems with neglected SOC.
\end{itemize}

Note that the results in the setting with SOC neglected may be used in some boson systems, such as phonon, photon and magnon systems.

In each of Secs. \ref{trssoc},\ref{trsnosoc},\ref{notrssoc},\ref{notrsnosoc}, we display the results for hourglass band crossings according to the following convention which  organizes these results in a compact manner to be friendly to the user as far as possible:

Following the order of 230 space groups, for each space group one by one, once it can host possible hourglass band crossing, we thus give the following information. Firstly we show essential results:\\

\begin{itemize}
  \item We first display essential results for hourglass nodal line (printed in red and bold style, which must occur in high symmetry plane), and the information for this case include:
      \begin{enumerate}
        \item the space group number (SG);
        \item the hourglass-relevant R-X-B (X, connecting R and B, can host hourglass band crossing, as described in the main text)  with explicit coordinates in conventional unit cell basis (all the labels for k vectors follow the convention in the Bilbao server \cite{Bilbao} where the coordinates can also be found);
        \item the splitting pattern related  with the hourglass band crossing: Note that we do not list compatibility relations since we are only concerned with splitting pattern as described in the main text, and  illustrated in Sec. \ref{sp} by an example;
        \item the hourglass type to which the hourglass nodal line belongs, i.e. type-a,b,c,d or e;
      \end{enumerate}
  \item Then display essential results for hourglass band crossings in high symmetry lines (in red), for which case, the hourglass band crossing may be just a hourglass nodal point, or lie in a (several) nodal line(s) lying in a (several) high symmetry plane(s). We give the information as the above, but note that when the hourglass band crossing lies in a (several) nodal line(s), we also give the concrete high symmetry plane(s) containing the nodal line(s) just under R-X-B (different high symmetry planes are separated by $\mid$).
\end{itemize}

 Then we display non-essential results following the above order, i.e. first giving information of all nonessential hourglass nodal lines (in black and bold style) and then all hourglass band crossings in high symmetry lines (printed in black, which could be a hourglass nodal point or lie in a (several) nodal line(s)).

    \subsection{Compatibility relations and splitting pattern}\label{sp}
  The hourglass band structure is intimately related to spatial symmetries. Concretely,  each hourglass is associated with four compatibility relations: \\\\

  \begin{itemize}
    \item   a irreducible representation (irrep) corresponding to the higher energy level at R is reduced to two irreps in X;
    \item   a irrep corresponding to the lower energy level at R is reduced to two irreps in X;
    \item   a irrep corresponding to the higher energy level at B is reduced to two irreps in X;
    \item    a irrep corresponding to the lower energy level at B is reduced to two irreps in X.
  \end{itemize}

  Note that the what matters is the splitting patterns of bands in X from R or B, and the details of the irrep(s) at R or B are not relevant with the formation of any hourglass band crossing at all. Hence, we display the compatibility relations just giving the irreps in X, for example, for space group 220 in the presence of TRS and SOC, the high symmetry plane C ($u,u,w$) could host a nonessential hourglass nodal line, since we find that for N $(\frac{1}{2},\frac{1}{2},0)$-C, all possible compatibility relations \cite{Bilbao} are listed in the following:
  \begin{equation}\label{sm-1}
    \begin{array}{c}
    \mathrm{N}2,5(2)\rightarrow 2\mathrm{C}4(1),\\
    \mathrm{N}3,4(2)\rightarrow 2\mathrm{C}3(1),\\
    \end{array}
  \end{equation}
  where the first line means that, the irrep N2,5 at N should split to two irreps which are both C4 in C, and so on, $\mathrm{C}4(1)$ represents 1d irrep C4 at C and $\mathrm{N}2,5(2)$ means that the 1d irreps $\mathrm{N}2$ and $\mathrm{N}5$ are paired by TRS to form a 2d (co-)irrep, and the number in the parenthesis behind the irrep label denotes the dimension of the (co-)irrep, and so on.\\
  For GM $(0,0,0)$-C, all possible compatibility relations \cite{Bilbao} are:

  \begin{equation}\label{sm-2}
    \begin{array}{c}
    \mathrm{GM}6(2)\rightarrow \mathrm{C}3(1)\oplus\mathrm{C}4(1)\\
       \mathrm{GM}7(2)\rightarrow \mathrm{C}3(1)\oplus\mathrm{C}4(1)\\
          \mathrm{GM}8(4)\rightarrow 2\mathrm{C}3(1)\oplus 2\mathrm{C}4(1)\\
    \end{array}
  \end{equation}
  Then for GM-C-N (corresponding to R-X-B in the general notation as described in the main text) in space group 220, all possible hourglass band crossings are nonessential and the concrete information of irreps are:\\
\begin{itemize}
\item 1.  \begin{itemize}
   \item  The higher energy level at GM belongs to one of GM6 and GM7, splitting to one C3 and one C4 in C;
   \item  The lower energy level at GM belongs to one of GM6 and GM7, splitting to one C3 and one C4 in C;
   \item  The higher energy level at N belongs to N2,5, splitting to two C4's;
   \item  The higher energy level at N belongs to N3,4, splitting to two  C3's,
 \end{itemize}
where it is easy to find that from GM to N, C3 and C4 in C are interchanged, which  just constitute  the irreps of the band crossing, contribute to a twofold degenerate hourglass nodal line. We use hereafter the following splitting pattern obtained from the compatibility relations as:
\begin{equation}\label{sm-ex}
  \begin{array}{c}
  \mathrm{C}3\oplus \mathrm{C}4,\\
  \mathrm{C}3\oplus \mathrm{C}4;\\
  2\mathrm{C}4,\\
  2\mathrm{C}3,
  \end{array}
\end{equation}
 corresponding to higher energy level at GM, lower energy level at GM, higher energy level at N and lowever energy level at N, respectively, to describe the splitting of the hourglass band crossing where the irreps at R or B are not specified at all,  otherwise there would be 2+2=4 possibilities corresponding to this kind of splitting.  This is because GM have 3 different irreps and two of them split into one C3 and one C4 in C. When the two energy levels in the hourglass structure at GM both belong to the same irrep, there are 2 different situations in total. When the two energy levels in the hourglass structure at GM belong to different irreps, there are 2!=2 possibilities in total. Thus the splitting pattern in Eq. \ref{sm-ex} corresponds to 4 different compatibility relations in total. Also, through interchanging band orderings at R or B would induce other splitting patterns:
\item 2.  As above, the second possibility of splitting pattern in GM-C-N after interchanging band ordering at N,  can be written as:
\begin{equation}\label{sm-ex-1}
  \begin{array}{c}
  \mathrm{C}3\oplus \mathrm{C}4,\\
  \mathrm{C}3\oplus \mathrm{C}4;\\
  2\mathrm{C}3,\\
  2\mathrm{C}4,
  \end{array}
\end{equation}
which is called equivalent to the splitting pattern in Eq. \ref{sm-ex}.
In next subsection, we show all such interchangings for all five types of hourglass band crossings where all possible equivalent hourglass structures are shown.

Owing to the above considerations, we only list the inequivalent splitting patterns for hourglass structures, otherwise there'll be too many redundant results.
\end{itemize}

Last but not least, we should point out that, once a combination of R-X-B could host hourglass band crossing, it is certain that B-X-R could as well and we just list one combination.
\begin{figure*}
  \centering
  \includegraphics[width=0.9\textwidth]{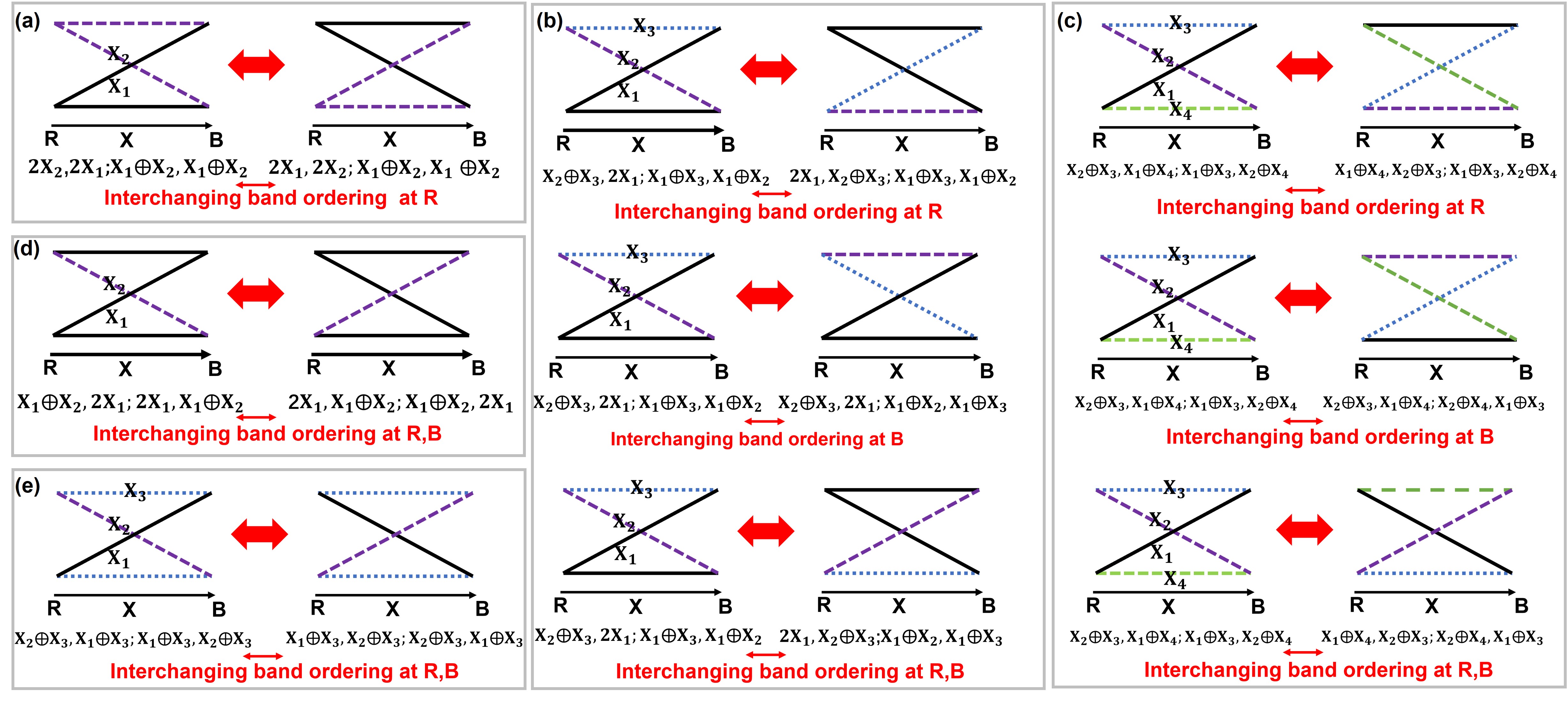}\\
  \caption{Schematic diagram of all equivalent hourglass structures in types-($a$-$e$): By interchanging the band ordering  at R or B, equivalent hourglass structures can be obtained. Below each hourglass structure is the splitting pattern as described in the main text.}\label{5types-sm}
\end{figure*}

\subsection{Equivalent hourglass band crossings}
Shown in Fig.\ref{5types-sm}, we display all possible situations that equivalent  hourglass structures are obtained by interchanging band ordering ar R or B. For each type, the left hourglass structure in each panel of Fig. \ref{5types-sm} corresponds to that as described in the main text.  We note that for both types $b$ and  $c$, they have four equivalent structures.

\section{Hourglass band crossings for materials with  TRS and significant SOC}\label{trssoc}
In this section, we list all possible positions in the BZ for 230 space groups by considering TRS and SOC.  Take Z-LD-GM in SG 4 as the example, namely, the first item in the table of this section. LD in SG 4 is obviously a high symmetry line and its coordinate in the conventional basis is $(0,v,0)$, can host an hourglass band crossing. Here, LD just corresponds to X in the abstract notation and connects Z and GM which correspond to R and B, respectively. We further find that such an hourglass band crossing is essential so Z-LD-GM is printed in red and also the hourglass band crossing in LD is an hourglass nodal point. For SG 7, in the row where the R-X-B is A-F-GM, we can find that A-F-GM is printed in red and bold style: this means that F can host an essential (denoted by red color) hourglass nodal line (denoted by bold style). Actually, we can directly know that F is a high symmetry plane since its coordinate is $(u,0,w)$ so that the hourglass band crossings in F must constitute a hourglass nodal line. Then we take U-A-Z in SG 28 as an example to show that A (it is a high symmetry line, so it is not in bold style) could host an essential (denoted by red color) hourglass band crossing which is further found to lie in a nodal line within M ($u,0,w$) as given just under U-A-Z in the table.  All in all, the above positions are all printed in red since they are all found to be essential while the high symmetry plane is printed in bold style.  We can also find some nonessential results which are printed in black. Besides, the above results are all of type a, which just corresponds to Fig. 1(a) of the main text. For next three settings, the corresponding tables can alo be read following the same way as described here.
\begin{center}\begin{table*}
\begin{tabular}{c|c|c|c}
\hline \hline
SG& R-X-B &  Splitting pattern in the hourglass band structure & Type  \\
\hline

4&{\color{red}{Z$(0,\frac{1}{2},0)$-LD$(0,v,0)$-GM$(0,0,0)$}}&$2\mathrm{LD}3(1),2\mathrm{LD}4(1);\mathrm{LD}3(1)\oplus\mathrm{LD}4(1),\mathrm{LD}3(1)\oplus\mathrm{LD}4(1)$&Type a\\ \hline
4&{\color{red}{E$(\frac{1}{2},\frac{1}{2},\frac{1}{2})$-U$(\frac{1}{2},v,\frac{1}{2})$-A$(\frac{1}{2},0,\frac{1}{2})$}}&$2\mathrm{U}3(1),2\mathrm{U}4(1);\mathrm{U}3(1)\oplus\mathrm{U}4(1),\mathrm{U}3(1)\oplus\mathrm{U}4(1)$&Type a\\ \hline
4&{\color{red}{D$(0,\frac{1}{2},\frac{1}{2})$-V$(0,v,\frac{1}{2})$-B$(0,0,\frac{1}{2})$}}&$2\mathrm{V}3(1),2\mathrm{V}4(1);\mathrm{V}3(1)\oplus\mathrm{V}4(1),\mathrm{V}3(1)\oplus\mathrm{V}4(1)$&Type a\\ \hline
4&{\color{red}{C$(\frac{1}{2},\frac{1}{2},0)$-W$(\frac{1}{2},v,0)$-Y$(\frac{1}{2},0,0)$}}&$2\mathrm{W}3(1),2\mathrm{W}4(1);\mathrm{W}3(1)\oplus\mathrm{W}4(1),\mathrm{W}3(1)\oplus\mathrm{W}4(1)$&Type a\\ \hline
7&{\bf\color{red}{A$(\frac{1}{2},0,\frac{1}{2})$-F$(u,0,w)$-GM$(0,0,0)$}}&$2\mathrm{F}3(1),2\mathrm{F}4(1);\mathrm{F}3(1)\oplus\mathrm{F}4(1),\mathrm{F}3(1)\oplus\mathrm{F}4(1)$&Type a\\ \hline
7&{\bf\color{red}{A$(\frac{1}{2},0,\frac{1}{2})$-F$(u,0,w)$-Y$(\frac{1}{2},0,0)$}}&$2\mathrm{F}3(1),2\mathrm{F}4(1);\mathrm{F}3(1)\oplus\mathrm{F}4(1),\mathrm{F}3(1)\oplus\mathrm{F}4(1)$&Type a\\ \hline
7&{\bf\color{red}{B$(0,0,\frac{1}{2})$-F$(u,0,w)$-GM$(0,0,0)$}}&$2\mathrm{F}3(1),2\mathrm{F}4(1);\mathrm{F}3(1)\oplus\mathrm{F}4(1),\mathrm{F}3(1)\oplus\mathrm{F}4(1)$&Type a\\
\hline
\hline \end{tabular}\end{table*}\end{center}

\newpage

\begin{center}\begin{table*}(Continued from previous table: Time-reversal symmetric, considering SOC)\\\begin{tabular}{c|c|c|c}
\hline \hline
SG& R-X-B & Splitting pattern in the  hourglss band structure & Type  \\
\hline

28&{\bf\color{red}{U$(\frac{1}{2},0,\frac{1}{2})$-M$(u,0,w)$-Z$(0,0,\frac{1}{2})$}}&$2\mathrm{M}3(1),2\mathrm{M}4(1);\mathrm{M}3(1)\oplus\mathrm{M}4(1),\mathrm{M}3(1)\oplus\mathrm{M}4(1)$&Type a\\ \hline
28&{\bf\color{red}{X$(\frac{1}{2},0,0)$-M$(u,0,w)$-GM$(0,0,0)$}}&$2\mathrm{M}3(1),2\mathrm{M}4(1);\mathrm{M}3(1)\oplus\mathrm{M}4(1),\mathrm{M}3(1)\oplus\mathrm{M}4(1)$&Type a\\ \hline
28&{\bf\color{red}{X$(\frac{1}{2},0,0)$-M$(u,0,w)$-LD$(0,0,w)$}}&$2\mathrm{M}3(1),2\mathrm{M}4(1);\mathrm{M}3(1)\oplus\mathrm{M}4(1),\mathrm{M}3(1)\oplus\mathrm{M}4(1)$&Type a\\ \hline
28&{\bf\color{red}{X$(\frac{1}{2},0,0)$-M$(u,0,w)$-Z$(0,0,\frac{1}{2})$}}&$2\mathrm{M}3(1),2\mathrm{M}4(1);\mathrm{M}3(1)\oplus\mathrm{M}4(1),\mathrm{M}3(1)\oplus\mathrm{M}4(1)$&Type a\\ \hline
28&{\bf\color{red}{R$(\frac{1}{2},\frac{1}{2},\frac{1}{2})$-N$(u,\frac{1}{2},w)$-H$(0,\frac{1}{2},w)$}}&$2\mathrm{N}3(1),2\mathrm{N}4(1);\mathrm{N}3(1)\oplus\mathrm{N}4(1),\mathrm{N}3(1)\oplus\mathrm{N}4(1)$&Type a\\ \hline
28&{\bf\color{red}{R$(\frac{1}{2},\frac{1}{2},\frac{1}{2})$-N$(u,\frac{1}{2},w)$-T$(0,\frac{1}{2},\frac{1}{2})$}}&$2\mathrm{N}3(1),2\mathrm{N}4(1);\mathrm{N}3(1)\oplus\mathrm{N}4(1),\mathrm{N}3(1)\oplus\mathrm{N}4(1)$&Type a\\ \hline
28&{\bf\color{red}{R$(\frac{1}{2},\frac{1}{2},\frac{1}{2})$-N$(u,\frac{1}{2},w)$-Y$(0,\frac{1}{2},0)$}}&$2\mathrm{N}3(1),2\mathrm{N}4(1);\mathrm{N}3(1)\oplus\mathrm{N}4(1),\mathrm{N}3(1)\oplus\mathrm{N}4(1)$&Type a\\
\hline
\hline \end{tabular}\end{table*}\end{center}

\newpage

\begin{center}\begin{table*}(Continued from previous table: Time-reversal symmetric, considering SOC)\\\begin{tabular}{c|c|c|c}
\hline \hline
SG& R-X-B &   Splitting pattern in the  hourglss band structure & Type  \\
\hline

29&{\color{red}{\makecell{X$(\frac{1}{2},0,0)$-SM$(u,0,0)$-GM$(0,0,0)$\\M$(u,0,w)$}}}&$2\mathrm{SM}3(1),2\mathrm{SM}4(1);\mathrm{SM}3(1)\oplus\mathrm{SM}4(1),\mathrm{SM}3(1)\oplus\mathrm{SM}4(1)$&Type a\\ \hline
29&\makecell{U$(\frac{1}{2},0,\frac{1}{2})$-G$(\frac{1}{2},0,w)$-X$(\frac{1}{2},0,0)$\\L$(\frac{1}{2},v,w)$}&$2\mathrm{G}2(1),2\mathrm{G}5(1);\mathrm{G}2(1)\oplus\mathrm{G}5(1),\mathrm{G}2(1)\oplus\mathrm{G}5(1)$&Type a\\ \hline
29&\makecell{U$(\frac{1}{2},0,\frac{1}{2})$-G$(\frac{1}{2},0,w)$-X$(\frac{1}{2},0,0)$\\L$(\frac{1}{2},v,w)$}&$2\mathrm{G}3(1),2\mathrm{G}4(1);\mathrm{G}3(1)\oplus\mathrm{G}4(1),\mathrm{G}3(1)\oplus\mathrm{G}4(1)$&Type a\\ \hline
29&\makecell{R$(\frac{1}{2},\frac{1}{2},\frac{1}{2})$-Q$(\frac{1}{2},\frac{1}{2},w)$-S$(\frac{1}{2},\frac{1}{2},0)$\\L$(\frac{1}{2},v,w)$}&$2\mathrm{Q}2(1),2\mathrm{Q}5(1);\mathrm{Q}2(1)\oplus\mathrm{Q}5(1),\mathrm{Q}2(1)\oplus\mathrm{Q}5(1)$&Type a\\ \hline
29&\makecell{R$(\frac{1}{2},\frac{1}{2},\frac{1}{2})$-Q$(\frac{1}{2},\frac{1}{2},w)$-S$(\frac{1}{2},\frac{1}{2},0)$\\L$(\frac{1}{2},v,w)$}&$2\mathrm{Q}3(1),2\mathrm{Q}4(1);\mathrm{Q}3(1)\oplus\mathrm{Q}4(1),\mathrm{Q}3(1)\oplus\mathrm{Q}4(1)$&Type a\\ \hline
30&{\bf\color{red}{Y$(0,\frac{1}{2},0)$-K$(0,v,w)$-GM$(0,0,0)$}}&$2\mathrm{K}3(1),2\mathrm{K}4(1);\mathrm{K}3(1)\oplus\mathrm{K}4(1),\mathrm{K}3(1)\oplus\mathrm{K}4(1)$&Type a\\ \hline
30&{\bf\color{red}{Y$(0,\frac{1}{2},0)$-K$(0,v,w)$-LD$(0,0,w)$}}&$2\mathrm{K}3(1),2\mathrm{K}4(1);\mathrm{K}3(1)\oplus\mathrm{K}4(1),\mathrm{K}3(1)\oplus\mathrm{K}4(1)$&Type a\\
\hline
\hline \end{tabular}\end{table*}\end{center}

\newpage

\begin{center}\begin{table*}(Continued from previous table: Time-reversal symmetric, considering SOC)\\\begin{tabular}{c|c|c|c}
\hline \hline
SG& R-X-B & Splitting pattern in the  hourglss band structure & Type  \\
\hline

31&{\bf\color{red}{S$(\frac{1}{2},\frac{1}{2},0)$-N$(u,\frac{1}{2},w)$-Y$(0,\frac{1}{2},0)$}}&$2\mathrm{N}4(1),2\mathrm{N}3(1);\mathrm{N}3(1)\oplus\mathrm{N}4(1),\mathrm{N}3(1)\oplus\mathrm{N}4(1)$&Type a\\ \hline
31&{\bf\color{red}{S$(\frac{1}{2},\frac{1}{2},0)$-N$(u,\frac{1}{2},w)$-E$(u,\frac{1}{2},\frac{1}{2})$}}&$2\mathrm{N}4(1),2\mathrm{N}3(1);\mathrm{N}3(1)\oplus\mathrm{N}4(1),\mathrm{N}3(1)\oplus\mathrm{N}4(1)$&Type a\\ \hline
31&{\color{red}{\makecell{S$(\frac{1}{2},\frac{1}{2},0)$-C$(u,\frac{1}{2},0)$-Y$(0,\frac{1}{2},0)$\\N$(u,\frac{1}{2},w)$}}}&$2\mathrm{C}4(1),2\mathrm{C}3(1);\mathrm{C}3(1)\oplus\mathrm{C}4(1),\mathrm{C}3(1)\oplus\mathrm{C}4(1)$&Type a\\ \hline
31&{\color{red}{\makecell{X$(\frac{1}{2},0,0)$-SM$(u,0,0)$-GM$(0,0,0)$\\M$(u,0,w)$}}}&$2\mathrm{SM}4(1),2\mathrm{SM}3(1);\mathrm{SM}3(1)\oplus\mathrm{SM}4(1),\mathrm{SM}3(1)\oplus\mathrm{SM}4(1)$&Type a\\ \hline
31&{\color{red}{\makecell{U$(\frac{1}{2},0,\frac{1}{2})$-G$(\frac{1}{2},0,w)$-X$(\frac{1}{2},0,0)$\\M$(u,0,w)$}}}&$\mathrm{G}2(1)\oplus\mathrm{G}4(1),\mathrm{G}3(1)\oplus\mathrm{G}5(1);\mathrm{G}3(1)\oplus\mathrm{G}4(1),\mathrm{G}2(1)\oplus\mathrm{G}5(1)$&Type c\\ \hline
31&{\color{red}{\makecell{R$(\frac{1}{2},\frac{1}{2},\frac{1}{2})$-Q$(\frac{1}{2},\frac{1}{2},w)$-S$(\frac{1}{2},\frac{1}{2},0)$\\N$(u,\frac{1}{2},w)$}}}&$\mathrm{Q}2(1)\oplus\mathrm{Q}4(1),\mathrm{Q}3(1)\oplus\mathrm{Q}5(1);\mathrm{Q}3(1)\oplus\mathrm{Q}4(1),\mathrm{Q}2(1)\oplus\mathrm{Q}5(1)$&Type c\\ \hline
32&{\bf\color{red}{T$(0,\frac{1}{2},\frac{1}{2})$-K$(0,v,w)$-GM$(0,0,0)$}}&$2\mathrm{K}3(1),2\mathrm{K}4(1);\mathrm{K}3(1)\oplus\mathrm{K}4(1),\mathrm{K}3(1)\oplus\mathrm{K}4(1)$&Type a\\
\hline
\hline \end{tabular}\end{table*}\end{center}

\newpage

\begin{center}\begin{table*}(Continued from previous table: Time-reversal symmetric, considering SOC)\\\begin{tabular}{c|c|c|c}
\hline \hline
SG& R-X-B & Splitting pattern in the  hourglss band structure & Type  \\
\hline

33&{\bf\color{red}{U$(\frac{1}{2},0,\frac{1}{2})$-M$(u,0,w)$-LD$(0,0,w)$}}&$2\mathrm{M}3(1),2\mathrm{M}4(1);\mathrm{M}3(1)\oplus\mathrm{M}4(1),\mathrm{M}3(1)\oplus\mathrm{M}4(1)$&Type a\\ \hline
33&{\bf\color{red}{X$(\frac{1}{2},0,0)$-M$(u,0,w)$-GM$(0,0,0)$}}&$2\mathrm{M}3(1),2\mathrm{M}4(1);\mathrm{M}3(1)\oplus\mathrm{M}4(1),\mathrm{M}3(1)\oplus\mathrm{M}4(1)$&Type a\\ \hline
33&{\bf\color{red}{X$(\frac{1}{2},0,0)$-M$(u,0,w)$-LD$(0,0,w)$}}&$2\mathrm{M}3(1),2\mathrm{M}4(1);\mathrm{M}3(1)\oplus\mathrm{M}4(1),\mathrm{M}3(1)\oplus\mathrm{M}4(1)$&Type a\\ \hline
33&{\bf\color{red}{A$(u,0,\frac{1}{2})$-M$(u,0,w)$-GM$(0,0,0)$}}&$2\mathrm{M}3(1),2\mathrm{M}4(1);\mathrm{M}3(1)\oplus\mathrm{M}4(1),\mathrm{M}3(1)\oplus\mathrm{M}4(1)$&Type a\\ \hline
33&{\bf\color{red}{A$(u,0,\frac{1}{2})$-M$(u,0,w)$-LD$(0,0,w)$}}&$2\mathrm{M}3(1),2\mathrm{M}4(1);\mathrm{M}3(1)\oplus\mathrm{M}4(1),\mathrm{M}3(1)\oplus\mathrm{M}4(1)$&Type a\\ \hline
33&{\color{red}{\makecell{Y$(0,\frac{1}{2},0)$-DT$(0,v,0)$-GM$(0,0,0)$\\K$(0,v,w)$}}}&$2\mathrm{DT}3(1),2\mathrm{DT}4(1);\mathrm{DT}3(1)\oplus\mathrm{DT}4(1),\mathrm{DT}3(1)\oplus\mathrm{DT}4(1)$&Type a\\ \hline
33&{\color{red}{\makecell{X$(\frac{1}{2},0,0)$-SM$(u,0,0)$-GM$(0,0,0)$\\M$(u,0,w)$}}}&$2\mathrm{SM}3(1),2\mathrm{SM}4(1);\mathrm{SM}3(1)\oplus\mathrm{SM}4(1),\mathrm{SM}3(1)\oplus\mathrm{SM}4(1)$&Type a\\
\hline
\hline \end{tabular}\end{table*}\end{center}

\newpage

\begin{center}\begin{table*}(Continued from previous table: Time-reversal symmetric, considering SOC)\\\begin{tabular}{c|c|c|c}
\hline \hline
SG& R-X-B &   Splitting pattern in the  hourglss band structure & Type  \\
\hline

34&{\color{red}{\makecell{T$(0,\frac{1}{2},\frac{1}{2})$-H$(0,\frac{1}{2},w)$-Y$(0,\frac{1}{2},0)$\\K$(0,v,w)$$\mid$N$(u,\frac{1}{2},w)$}}}&$\mathrm{H}3(1)\oplus\mathrm{H}4(1),\mathrm{H}2(1)\oplus\mathrm{H}5(1);\mathrm{H}2(1)\oplus\mathrm{H}3(1),\mathrm{H}4(1)\oplus\mathrm{H}5(1)$&Type c\\ \hline
36&{\bf\color{red}{B$(0,v,\frac{1}{2})$-K$(0,v,w)$-GM$(0,0,0)$}}&$2\mathrm{K}3(1),2\mathrm{K}4(1);\mathrm{K}3(1)\oplus\mathrm{K}4(1),\mathrm{K}3(1)\oplus\mathrm{K}4(1)$&Type a\\ \hline
36&{\bf\color{red}{B$(0,v,\frac{1}{2})$-K$(0,v,w)$-H$(0,1,w)$}}&$2\mathrm{K}3(1),2\mathrm{K}4(1);\mathrm{K}3(1)\oplus\mathrm{K}4(1),\mathrm{K}3(1)\oplus\mathrm{K}4(1)$&Type a\\ \hline
36&{\bf\color{red}{B$(0,v,\frac{1}{2})$-K$(0,v,w)$-LD$(0,0,w)$}}&$2\mathrm{K}3(1),2\mathrm{K}4(1);\mathrm{K}3(1)\oplus\mathrm{K}4(1),\mathrm{K}3(1)\oplus\mathrm{K}4(1)$&Type a\\ \hline
36&{\bf\color{red}{B$(0,v,\frac{1}{2})$-K$(0,v,w)$-Y$(0,1,0)$}}&$2\mathrm{K}3(1),2\mathrm{K}4(1);\mathrm{K}3(1)\oplus\mathrm{K}4(1),\mathrm{K}3(1)\oplus\mathrm{K}4(1)$&Type a\\ \hline
36&{\color{red}{R$(\frac{1}{2},\frac{1}{2},\frac{1}{2})$-D$(\frac{1}{2},\frac{1}{2},w)$-S$(\frac{1}{2},\frac{1}{2},0)$}}&$2\mathrm{D}3(1),2\mathrm{D}4(1);\mathrm{D}3(1)\oplus\mathrm{D}4(1),\mathrm{D}3(1)\oplus\mathrm{D}4(1)$&Type a\\ \hline
39&{\bf\color{red}{S$(0,\frac{1}{2},-\frac{1}{2})$-P$(0,v,w)$-GM$(0,0,0)$}}&$2\mathrm{P}3(1),2\mathrm{P}4(1);\mathrm{P}3(1)\oplus\mathrm{P}4(1),\mathrm{P}3(1)\oplus\mathrm{P}4(1)$&Type a\\
\hline
\hline \end{tabular}\end{table*}\end{center}

\newpage

\begin{center}\begin{table*}(Continued from previous table: Time-reversal symmetric, considering SOC)\\\begin{tabular}{c|c|c|c}
\hline \hline
SG& R-X-B & Splitting pattern in the  hourglss band structure & Type  \\
\hline

43&{\bf\color{red}{Y$(0,1,0)$-E$(0,v,w)$-B$(0,v,1)$}}&$2\mathrm{E}3(1),2\mathrm{E}4(1);\mathrm{E}3(1)\oplus\mathrm{E}4(1),\mathrm{E}3(1)\oplus\mathrm{E}4(1)$&Type a\\ \hline
43&{\bf\color{red}{T$(1,0,0)$-J$(u,0,w)$-GM$(0,0,0)$}}&$2\mathrm{J}3(1),2\mathrm{J}4(1);\mathrm{J}3(1)\oplus\mathrm{J}4(1),\mathrm{J}3(1)\oplus\mathrm{J}4(1)$&Type a\\ \hline
43&{\bf\color{red}{T$(1,0,0)$-J$(u,0,w)$-LD$(0,0,w)$}}&$2\mathrm{J}3(1),2\mathrm{J}4(1);\mathrm{J}3(1)\oplus\mathrm{J}4(1),\mathrm{J}3(1)\oplus\mathrm{J}4(1)$&Type a\\ \hline
43&{\bf\color{red}{T$(1,0,0)$-J$(u,0,w)$-Y$(1,0,1)$}}&$2\mathrm{J}3(1),2\mathrm{J}4(1);\mathrm{J}3(1)\oplus\mathrm{J}4(1),\mathrm{J}3(1)\oplus\mathrm{J}4(1)$&Type a\\ \hline
43&{\bf\color{red}{T$(1,0,0)$-J$(u,0,w)$-A$(u,0,1)$}}&$2\mathrm{J}3(1),2\mathrm{J}4(1);\mathrm{J}3(1)\oplus\mathrm{J}4(1),\mathrm{J}3(1)\oplus\mathrm{J}4(1)$&Type a\\ \hline
43&{\color{red}{\makecell{Y$(0,1,0)$-DT$(0,v,0)$-GM$(0,0,0)$\\E$(0,v,w)$}}}&$2\mathrm{DT}3(1),2\mathrm{DT}4(1);\mathrm{DT}3(1)\oplus\mathrm{DT}4(1),\mathrm{DT}3(1)\oplus\mathrm{DT}4(1)$&Type a\\ \hline
43&{\color{red}{\makecell{T$(1,0,0)$-SM$(u,0,0)$-GM$(0,0,0)$\\J$(u,0,w)$}}}&$2\mathrm{SM}3(1),2\mathrm{SM}4(1);\mathrm{SM}3(1)\oplus\mathrm{SM}4(1),\mathrm{SM}3(1)\oplus\mathrm{SM}4(1)$&Type a\\
\hline
\hline \end{tabular}\end{table*}\end{center}

\newpage

\begin{center}\begin{table*}(Continued from previous table: Time-reversal symmetric, considering SOC)\\\begin{tabular}{c|c|c|c}
\hline \hline
SG& R-X-B & Splitting pattern in the  hourglss band structure & Type  \\
\hline

60&{\bf\color{red}{U$(\frac{1}{2},0,\frac{1}{2})$-W$(u,v,\frac{1}{2})$-Z$(0,0,\frac{1}{2})$}}&$2\mathrm{W}4,4(2),2\mathrm{W}3,3(2);\mathrm{W}3,3(2)\oplus\mathrm{W}4,4(2),\mathrm{W}3,3(2)\oplus\mathrm{W}4,4(2)$&Type a\\ \hline
60&{\bf\color{red}{U$(\frac{1}{2},0,\frac{1}{2})$-W$(u,v,\frac{1}{2})$-B$(0,v,\frac{1}{2})$}}&$2\mathrm{W}4,4(2),2\mathrm{W}3,3(2);\mathrm{W}3,3(2)\oplus\mathrm{W}4,4(2),\mathrm{W}3,3(2)\oplus\mathrm{W}4,4(2)$&Type a\\ \hline
60&{\bf\color{red}{U$(\frac{1}{2},0,\frac{1}{2})$-W$(u,v,\frac{1}{2})$-E$(u,\frac{1}{2},\frac{1}{2})$}}&$2\mathrm{W}4,4(2),2\mathrm{W}3,3(2);\mathrm{W}3,3(2)\oplus\mathrm{W}4,4(2),\mathrm{W}3,3(2)\oplus\mathrm{W}4,4(2)$&Type a\\ \hline
60&{\color{red}{T$(0,\frac{1}{2},\frac{1}{2})$-H$(0,\frac{1}{2},w)$-Y$(0,\frac{1}{2},0)$}}&$2\mathrm{H}3,5(2),2\mathrm{H}2,4(2);\mathrm{H}2,4(2)\oplus\mathrm{H}3,5(2),\mathrm{H}2,4(2)\oplus\mathrm{H}3,5(2)$&Type a\\ \hline
60&{\color{red}{\makecell{U$(\frac{1}{2},0,\frac{1}{2})$-A$(u,0,\frac{1}{2})$-Z$(0,0,\frac{1}{2})$\\W$(u,v,\frac{1}{2})$}}}&$2\mathrm{A}2,5(2),2\mathrm{A}3,4(2);\mathrm{A}2,5(2)\oplus\mathrm{A}3,4(2),\mathrm{A}2,5(2)\oplus\mathrm{A}3,4(2)$&Type a\\ \hline
60&{\color{red}{\makecell{R$(\frac{1}{2},\frac{1}{2},\frac{1}{2})$-P$(\frac{1}{2},v,\frac{1}{2})$-U$(\frac{1}{2},0,\frac{1}{2})$\\L$(\frac{1}{2},v,w)$$\mid$W$(u,v,\frac{1}{2})$}}}&$\mathrm{P}2,2(2)\oplus\mathrm{P}3,3(2),\mathrm{P}4,4(2)\oplus\mathrm{P}5,5(2);\mathrm{P}3,3(2)\oplus\mathrm{P}4,4(2),\mathrm{P}2,2(2)\oplus\mathrm{P}5,5(2)$&Type c\\ \hline
61&{\bf\color{red}{S$(\frac{1}{2},\frac{1}{2},0)$-L$(\frac{1}{2},v,w)$-U$(\frac{1}{2},0,\frac{1}{2})$}}&$2\mathrm{L}3,3(2),2\mathrm{L}4,4(2);\mathrm{L}3,3(2)\oplus\mathrm{L}4,4(2),\mathrm{L}3,3(2)\oplus\mathrm{L}4,4(2)$&Type a\\
\hline
\hline \end{tabular}\end{table*}\end{center}

\newpage

\begin{center}\begin{table*}(Continued from previous table: Time-reversal symmetric, considering SOC)\\\begin{tabular}{c|c|c|c}
\hline \hline
SG& R-X-B &    Splitting pattern in the  hourglss band structure & Type  \\
\hline

93&{\color{red}{GM$(0,0,0)$-LD$(0,0,w)$-Z$(0,0,\frac{1}{2})$}}&$\mathrm{LD}5(1)\oplus\mathrm{LD}7(1),\mathrm{LD}6(1)\oplus\mathrm{LD}8(1);\mathrm{LD}6(1)\oplus\mathrm{LD}7(1),\mathrm{LD}5(1)\oplus\mathrm{LD}8(1)$&Type c\\ \hline
93&{\color{red}{A$(\frac{1}{2},\frac{1}{2},\frac{1}{2})$-V$(\frac{1}{2},\frac{1}{2},w)$-M$(\frac{1}{2},\frac{1}{2},0)$}}&$\mathrm{V}5(1)\oplus\mathrm{V}8(1),\mathrm{V}6(1)\oplus\mathrm{V}7(1);\mathrm{V}6(1)\oplus\mathrm{V}8(1),\mathrm{V}5(1)\oplus\mathrm{V}7(1)$&Type c\\ \hline
94&{\color{red}{X$(0,\frac{1}{2},0)$-DT$(0,v,0)$-GM$(0,0,0)$}}&$2\mathrm{DT}3(1),2\mathrm{DT}4(1);\mathrm{DT}3(1)\oplus\mathrm{DT}4(1),\mathrm{DT}3(1)\oplus\mathrm{DT}4(1)$&Type a\\ \hline
94&{\color{red}{R$(0,\frac{1}{2},\frac{1}{2})$-U$(0,v,\frac{1}{2})$-Z$(0,0,\frac{1}{2})$}}&$2\mathrm{U}3(1),2\mathrm{U}4(1);\mathrm{U}3(1)\oplus\mathrm{U}4(1),\mathrm{U}3(1)\oplus\mathrm{U}4(1)$&Type a\\ \hline
94&{\color{red}{GM$(0,0,0)$-LD$(0,0,w)$-Z$(0,0,\frac{1}{2})$}}&$\mathrm{LD}5(1)\oplus\mathrm{LD}7(1),\mathrm{LD}6(1)\oplus\mathrm{LD}8(1);\mathrm{LD}6(1)\oplus\mathrm{LD}7(1),\mathrm{LD}5(1)\oplus\mathrm{LD}8(1)$&Type c\\ \hline
95&{\color{red}{R$(0,\frac{1}{2},\frac{1}{2})$-W$(0,\frac{1}{2},w)$-X$(0,\frac{1}{2},0)$}}&$2\mathrm{W}3(1),2\mathrm{W}4(1);\mathrm{W}3(1)\oplus\mathrm{W}4(1),\mathrm{W}3(1)\oplus\mathrm{W}4(1)$&Type a\\ \hline
96&{\color{red}{X$(0,\frac{1}{2},0)$-DT$(0,v,0)$-GM$(0,0,0)$}}&$2\mathrm{DT}3(1),2\mathrm{DT}4(1);\mathrm{DT}3(1)\oplus\mathrm{DT}4(1),\mathrm{DT}3(1)\oplus\mathrm{DT}4(1)$&Type a\\
\hline
\hline \end{tabular}\end{table*}\end{center}

\newpage

\begin{center}\begin{table*}(Continued from previous table: Time-reversal symmetric, considering SOC)\\\begin{tabular}{c|c|c|c}
\hline \hline
SG& R-X-B &   Splitting pattern in the  hourglss band structure & Type  \\
\hline

104&{\color{red}{\makecell{R$(0,\frac{1}{2},\frac{1}{2})$-W$(0,\frac{1}{2},w)$-X$(0,\frac{1}{2},0)$\\B$(0,v,w)$$\mid$F$(u,\frac{1}{2},w)$}}}&$\mathrm{W}3(1)\oplus\mathrm{W}4(1),\mathrm{W}2(1)\oplus\mathrm{W}5(1);\mathrm{W}2(1)\oplus\mathrm{W}3(1),\mathrm{W}4(1)\oplus\mathrm{W}5(1)$&Type c\\ \hline
106&{\bf\color{red}{R$(0,\frac{1}{2},\frac{1}{2})$-B$(0,v,w)$-GM$(0,0,0)$}}&$2\mathrm{B}3(1),2\mathrm{B}4(1);\mathrm{B}3(1)\oplus\mathrm{B}4(1),\mathrm{B}3(1)\oplus\mathrm{B}4(1)$&Type a\\ \hline
106&{\bf\color{red}{R$(0,\frac{1}{2},\frac{1}{2})$-B$(0,v,w)$-LD$(0,0,w)$}}&$2\mathrm{B}3(1),2\mathrm{B}4(1);\mathrm{B}3(1)\oplus\mathrm{B}4(1),\mathrm{B}3(1)\oplus\mathrm{B}4(1)$&Type a\\ \hline
106&{\bf\color{red}{X$(0,\frac{1}{2},0)$-B$(0,v,w)$-GM$(0,0,0)$}}&$2\mathrm{B}3(1),2\mathrm{B}4(1);\mathrm{B}3(1)\oplus\mathrm{B}4(1),\mathrm{B}3(1)\oplus\mathrm{B}4(1)$&Type a\\ \hline
106&{\bf\color{red}{X$(0,\frac{1}{2},0)$-B$(0,v,w)$-LD$(0,0,w)$}}&$2\mathrm{B}3(1),2\mathrm{B}4(1);\mathrm{B}3(1)\oplus\mathrm{B}4(1),\mathrm{B}3(1)\oplus\mathrm{B}4(1)$&Type a\\ \hline
106&{\color{red}{A$(\frac{1}{2},\frac{1}{2},\frac{1}{2})$-V$(\frac{1}{2},\frac{1}{2},w)$-M$(\frac{1}{2},\frac{1}{2},0)$}}&$2\mathrm{V}6(2),2\mathrm{V}7(2);\mathrm{V}6(2)\oplus\mathrm{V}7(2),\mathrm{V}6(2)\oplus\mathrm{V}7(2)$&Type a\\ \hline
106&{\color{red}{\makecell{X$(0,\frac{1}{2},0)$-DT$(0,v,0)$-GM$(0,0,0)$\\B$(0,v,w)$}}}&$2\mathrm{DT}3(1),2\mathrm{DT}4(1);\mathrm{DT}3(1)\oplus\mathrm{DT}4(1),\mathrm{DT}3(1)\oplus\mathrm{DT}4(1)$&Type a\\
\hline
\hline \end{tabular}\end{table*}\end{center}

\newpage

\begin{center}\begin{table*}(Continued from previous table: Time-reversal symmetric, considering SOC)\\\begin{tabular}{c|c|c|c}
\hline \hline
SG& R-X-B & Splitting pattern in the  hourglss band structure & Type  \\
\hline

118&{\bf\color{red}{X$(0,\frac{1}{2},0)$-B$(0,v,w)$-GM$(0,0,0)$}}&$2\mathrm{B}3(1),2\mathrm{B}4(1);\mathrm{B}3(1)\oplus\mathrm{B}4(1),\mathrm{B}3(1)\oplus\mathrm{B}4(1)$&Type a\\ \hline
118&{\bf\color{red}{X$(0,\frac{1}{2},0)$-B$(0,v,w)$-LD$(0,0,w)$}}&$2\mathrm{B}3(1),2\mathrm{B}4(1);\mathrm{B}3(1)\oplus\mathrm{B}4(1),\mathrm{B}3(1)\oplus\mathrm{B}4(1)$&Type a\\ \hline
118&{\bf\color{red}{X$(0,\frac{1}{2},0)$-B$(0,v,w)$-R$(0,\frac{1}{2},\frac{1}{2})$}}&$2\mathrm{B}3(1),2\mathrm{B}4(1);\mathrm{B}3(1)\oplus\mathrm{B}4(1),\mathrm{B}3(1)\oplus\mathrm{B}4(1)$&Type a\\ \hline
118&{\bf\color{red}{X$(0,\frac{1}{2},0)$-B$(0,v,w)$-U$(0,v,\frac{1}{2})$}}&$2\mathrm{B}3(1),2\mathrm{B}4(1);\mathrm{B}3(1)\oplus\mathrm{B}4(1),\mathrm{B}3(1)\oplus\mathrm{B}4(1)$&Type a\\ \hline
118&{\bf\color{red}{R$(0,\frac{1}{2},\frac{1}{2})$-F$(u,\frac{1}{2},w)$-A$(\frac{1}{2},\frac{1}{2},\frac{1}{2})$}}&$2\mathrm{F}3(1),2\mathrm{F}4(1);\mathrm{F}3(1)\oplus\mathrm{F}4(1),\mathrm{F}3(1)\oplus\mathrm{F}4(1)$&Type a\\ \hline
118&{\bf\color{red}{R$(0,\frac{1}{2},\frac{1}{2})$-F$(u,\frac{1}{2},w)$-V$(\frac{1}{2},\frac{1}{2},w)$}}&$2\mathrm{F}3(1),2\mathrm{F}4(1);\mathrm{F}3(1)\oplus\mathrm{F}4(1),\mathrm{F}3(1)\oplus\mathrm{F}4(1)$&Type a\\ \hline
118&{\bf\color{red}{R$(0,\frac{1}{2},\frac{1}{2})$-F$(u,\frac{1}{2},w)$-X$(0,\frac{1}{2},0)$}}&$2\mathrm{F}3(1),2\mathrm{F}4(1);\mathrm{F}3(1)\oplus\mathrm{F}4(1),\mathrm{F}3(1)\oplus\mathrm{F}4(1)$&Type a\\
\hline
\hline \end{tabular}\end{table*}\end{center}

\newpage

\begin{center}\begin{table*}(Continued from previous table: Time-reversal symmetric, considering SOC)\\\begin{tabular}{c|c|c|c}
\hline \hline
SG& R-X-B &   Splitting pattern in the  hourglss band structure & Type  \\
\hline

159&{\color{red}{\makecell{L$(\frac{1}{2},0,\frac{1}{2})$-U$(\frac{1}{2},0,w)$-M$(\frac{1}{2},0,0)$\\C$(u,u,w)$}}}&$2\mathrm{U}3(1),2\mathrm{U}4(1);\mathrm{U}3(1)\oplus\mathrm{U}4(1),\mathrm{U}3(1)\oplus\mathrm{U}4(1)$&Type a\\ \hline
159&\textbf{A$(0,0,\frac{1}{2})$-C$(u,u,w)$-DT$(0,0,w)$}&$2\mathrm{C}3(1),2\mathrm{C}4(1);\mathrm{C}3(1)\oplus\mathrm{C}4(1),\mathrm{C}3(1)\oplus\mathrm{C}4(1)$&Type a\\ \hline
159&\textbf{A$(0,0,\frac{1}{2})$-C$(u,u,w)$-GM$(0,0,0)$}&$2\mathrm{C}3(1),2\mathrm{C}4(1);\mathrm{C}3(1)\oplus\mathrm{C}4(1),\mathrm{C}3(1)\oplus\mathrm{C}4(1)$&Type a\\ \hline
159&\textbf{A$(0,0,\frac{1}{2})$-C$(u,u,w)$-H$(\frac{1}{3},\frac{1}{3},\frac{1}{2})$}&$2\mathrm{C}3(1),2\mathrm{C}4(1);\mathrm{C}3(1)\oplus\mathrm{C}4(1),\mathrm{C}3(1)\oplus\mathrm{C}4(1)$&Type a\\ \hline
159&\textbf{A$(0,0,\frac{1}{2})$-C$(u,u,w)$-HA$(-\frac{1}{3},-\frac{1}{3},-\frac{1}{2})$}&$2\mathrm{C}3(1),2\mathrm{C}4(1);\mathrm{C}3(1)\oplus\mathrm{C}4(1),\mathrm{C}3(1)\oplus\mathrm{C}4(1)$&Type a\\ \hline
159&\textbf{A$(0,0,\frac{1}{2})$-C$(u,u,w)$-K$(\frac{1}{3},\frac{1}{3},0)$}&$2\mathrm{C}3(1),2\mathrm{C}4(1);\mathrm{C}3(1)\oplus\mathrm{C}4(1),\mathrm{C}3(1)\oplus\mathrm{C}4(1)$&Type a\\ \hline
159&\textbf{A$(0,0,\frac{1}{2})$-C$(u,u,w)$-KA$(-\frac{1}{3},-\frac{1}{3},0)$}&$2\mathrm{C}3(1),2\mathrm{C}4(1);\mathrm{C}3(1)\oplus\mathrm{C}4(1),\mathrm{C}3(1)\oplus\mathrm{C}4(1)$&Type a\\
\hline
\hline \end{tabular}\end{table*}\end{center}

\newpage

\begin{center}\begin{table*}(Continued from previous table: Time-reversal symmetric, considering SOC)\\\begin{tabular}{c|c|c|c}
\hline \hline
SG& R-X-B & Splitting pattern in the  hourglss band structure & Type  \\
\hline

185&{\bf\color{red}{Q$(u,u,\frac{1}{2})$-C$(u,u,w)$-M$(\frac{1}{2},\frac{1}{2},0)$}}&$2\mathrm{C}3(1),2\mathrm{C}4(1);\mathrm{C}3(1)\oplus\mathrm{C}4(1),\mathrm{C}3(1)\oplus\mathrm{C}4(1)$&Type a\\ \hline
185&{\bf\color{red}{Q$(u,u,\frac{1}{2})$-C$(u,u,w)$-U$(\frac{1}{2},\frac{1}{2},w)$}}&$2\mathrm{C}3(1),2\mathrm{C}4(1);\mathrm{C}3(1)\oplus\mathrm{C}4(1),\mathrm{C}3(1)\oplus\mathrm{C}4(1)$&Type a\\ \hline
185&\textbf{H$(\frac{1}{3},\frac{1}{3},\frac{1}{2})$-C$(u,u,w)$-DT$(0,0,w)$}&$2\mathrm{C}3(1),2\mathrm{C}4(1);\mathrm{C}3(1)\oplus\mathrm{C}4(1),\mathrm{C}3(1)\oplus\mathrm{C}4(1)$&Type a\\ \hline
185&\textbf{H$(\frac{1}{3},\frac{1}{3},\frac{1}{2})$-C$(u,u,w)$-GM$(0,0,0)$}&$2\mathrm{C}3(1),2\mathrm{C}4(1);\mathrm{C}3(1)\oplus\mathrm{C}4(1),\mathrm{C}3(1)\oplus\mathrm{C}4(1)$&Type a\\ \hline
185&\textbf{H$(\frac{1}{3},\frac{1}{3},\frac{1}{2})$-C$(u,u,w)$-K$(\frac{1}{3},\frac{1}{3},0)$}&$2\mathrm{C}3(1),2\mathrm{C}4(1);\mathrm{C}3(1)\oplus\mathrm{C}4(1),\mathrm{C}3(1)\oplus\mathrm{C}4(1)$&Type a\\ \hline
185&\textbf{H$(\frac{1}{3},\frac{1}{3},\frac{1}{2})$-C$(u,u,w)$-P$(\frac{1}{3},\frac{1}{3},w)$}&$2\mathrm{C}3(1),2\mathrm{C}4(1);\mathrm{C}3(1)\oplus\mathrm{C}4(1),\mathrm{C}3(1)\oplus\mathrm{C}4(1)$&Type a\\ \hline
185&\textbf{H$(\frac{1}{3},\frac{1}{3},\frac{1}{2})$-C$(u,u,w)$-M$(\frac{1}{2},\frac{1}{2},0)$}&$2\mathrm{C}3(1),2\mathrm{C}4(1);\mathrm{C}3(1)\oplus\mathrm{C}4(1),\mathrm{C}3(1)\oplus\mathrm{C}4(1)$&Type a\\
\hline
\hline \end{tabular}\end{table*}\end{center}

\newpage

\begin{center}\begin{table*}(Continued from previous table: Time-reversal symmetric, considering SOC)\\\begin{tabular}{c|c|c|c}
\hline \hline
SG& R-X-B &  Splitting pattern in the  hourglss band structure & Type  \\
\hline

190&\textbf{A$(0,0,\frac{1}{2})$-C$(u,u,w)$-LD$(u,u,0)$}&$2\mathrm{C}4(1),2\mathrm{C}3(1);\mathrm{C}3(1)\oplus\mathrm{C}4(1),\mathrm{C}3(1)\oplus\mathrm{C}4(1)$&Type a\\ \hline
190&\textbf{A$(0,0,\frac{1}{2})$-C$(u,u,w)$-M$(\frac{1}{2},\frac{1}{2},0)$}&$2\mathrm{C}4(1),2\mathrm{C}3(1);\mathrm{C}3(1)\oplus\mathrm{C}4(1),\mathrm{C}3(1)\oplus\mathrm{C}4(1)$&Type a\\ \hline
190&\textbf{L$(\frac{1}{2},\frac{1}{2},\frac{1}{2})$-C$(u,u,w)$-H$(\frac{1}{3},\frac{1}{3},\frac{1}{2})$}&$2\mathrm{C}3(1),2\mathrm{C}4(1);\mathrm{C}3(1)\oplus\mathrm{C}4(1),\mathrm{C}3(1)\oplus\mathrm{C}4(1)$&Type a\\ \hline
190&\textbf{L$(\frac{1}{2},\frac{1}{2},\frac{1}{2})$-C$(u,u,w)$-HA$(-\frac{1}{3},-\frac{1}{3},-\frac{1}{2})$}&$2\mathrm{C}3(1),2\mathrm{C}4(1);\mathrm{C}3(1)\oplus\mathrm{C}4(1),\mathrm{C}3(1)\oplus\mathrm{C}4(1)$&Type a\\ \hline
190&\textbf{L$(\frac{1}{2},\frac{1}{2},\frac{1}{2})$-C$(u,u,w)$-DT$(0,0,w)$}&$2\mathrm{C}3(1),2\mathrm{C}4(1);\mathrm{C}3(1)\oplus\mathrm{C}4(1),\mathrm{C}3(1)\oplus\mathrm{C}4(1)$&Type a\\ \hline
190&\textbf{L$(\frac{1}{2},\frac{1}{2},\frac{1}{2})$-C$(u,u,w)$-P$(\frac{1}{3},\frac{1}{3},w)$}&$2\mathrm{C}3(1),2\mathrm{C}4(1);\mathrm{C}3(1)\oplus\mathrm{C}4(1),\mathrm{C}3(1)\oplus\mathrm{C}4(1)$&Type a\\ \hline
190&\textbf{L$(\frac{1}{2},\frac{1}{2},\frac{1}{2})$-C$(u,u,w)$-PA$(-\frac{1}{3},-\frac{1}{3},-w)$}&$2\mathrm{C}3(1),2\mathrm{C}4(1);\mathrm{C}3(1)\oplus\mathrm{C}4(1),\mathrm{C}3(1)\oplus\mathrm{C}4(1)$&Type a\\
\hline
\hline \end{tabular}\end{table*}\end{center}

\newpage

\section{Hourglass band crossings for materials with  TRS and negligible SOC}\label{trsnosoc}
In this section, we list all possible positions in the BZ for 230 space groups by considering TRS while neglecting SOC.
\begin{center}\begin{table*}\begin{tabular}{c|c|c|c}
\hline \hline
SG& R-X-B &  Splitting pattern in the hourglass band structure & Type  \\
\hline

29&{\bf\color{red}{P$(\frac{1}{2},v,\frac{1}{2})$-L$(\frac{1}{2},v,w)$-G$(\frac{1}{2},0,w)$}}&$2\mathrm{L}2(1),2\mathrm{L}1(1);\mathrm{L}1(1)\oplus\mathrm{L}2(1),\mathrm{L}1(1)\oplus\mathrm{L}2(1)$&Type a\\ \hline
29&{\bf\color{red}{P$(\frac{1}{2},v,\frac{1}{2})$-L$(\frac{1}{2},v,w)$-Q$(\frac{1}{2},\frac{1}{2},w)$}}&$2\mathrm{L}2(1),2\mathrm{L}1(1);\mathrm{L}1(1)\oplus\mathrm{L}2(1),\mathrm{L}1(1)\oplus\mathrm{L}2(1)$&Type a\\ \hline
29&{\bf\color{red}{P$(\frac{1}{2},v,\frac{1}{2})$-L$(\frac{1}{2},v,w)$-S$(\frac{1}{2},\frac{1}{2},0)$}}&$2\mathrm{L}2(1),2\mathrm{L}1(1);\mathrm{L}1(1)\oplus\mathrm{L}2(1),\mathrm{L}1(1)\oplus\mathrm{L}2(1)$&Type a\\ \hline
29&{\bf\color{red}{P$(\frac{1}{2},v,\frac{1}{2})$-L$(\frac{1}{2},v,w)$-X$(\frac{1}{2},0,0)$}}&$2\mathrm{L}2(1),2\mathrm{L}1(1);\mathrm{L}1(1)\oplus\mathrm{L}2(1),\mathrm{L}1(1)\oplus\mathrm{L}2(1)$&Type a\\ \hline
29&{\bf\color{red}{P$(\frac{1}{2},v,\frac{1}{2})$-L$(\frac{1}{2},v,w)$-D$(\frac{1}{2},v,0)$}}&$2\mathrm{L}2(1),2\mathrm{L}1(1);\mathrm{L}1(1)\oplus\mathrm{L}2(1),\mathrm{L}1(1)\oplus\mathrm{L}2(1)$&Type a\\ \hline
29&{\bf\color{red}{Z$(0,0,\frac{1}{2})$-M$(u,0,w)$-G$(\frac{1}{2},0,w)$}}&$2\mathrm{M}1(1),2\mathrm{M}2(1);\mathrm{M}1(1)\oplus\mathrm{M}2(1),\mathrm{M}1(1)\oplus\mathrm{M}2(1)$&Type a\\ \hline
29&{\bf\color{red}{Z$(0,0,\frac{1}{2})$-M$(u,0,w)$-X$(\frac{1}{2},0,0)$}}&$2\mathrm{M}1(1),2\mathrm{M}2(1);\mathrm{M}1(1)\oplus\mathrm{M}2(1),\mathrm{M}1(1)\oplus\mathrm{M}2(1)$&Type a\\
\hline
\hline \end{tabular}\end{table*}\end{center}

\newpage

\begin{center}\begin{table*}(Continued from previous table: Time-reversal symmetric, neglecting SOC)\\\begin{tabular}{c|c|c|c}
\hline \hline
SG& R-X-B &   Splitting pattern in the  hourglss band structure & Type  \\
\hline

52&{\color{red}{\makecell{R$(\frac{1}{2},\frac{1}{2},\frac{1}{2})$-P$(\frac{1}{2},v,\frac{1}{2})$-U$(\frac{1}{2},0,\frac{1}{2})$\\L$(\frac{1}{2},v,w)$}}}&$\mathrm{P}1(1)\oplus\mathrm{P}3(1),\mathrm{P}2(1)\oplus\mathrm{P}4(1);\mathrm{P}1(1)\oplus\mathrm{P}2(1),\mathrm{P}3(1)\oplus\mathrm{P}4(1)$&Type c\\ \hline
54&{\bf\color{red}{X$(\frac{1}{2},0,0)$-M$(u,0,w)$-Z$(0,0,\frac{1}{2})$}}&$2\mathrm{M}1(1),2\mathrm{M}2(1);\mathrm{M}1(1)\oplus\mathrm{M}2(1),\mathrm{M}1(1)\oplus\mathrm{M}2(1)$&Type a\\ \hline
54&{\bf\color{red}{X$(\frac{1}{2},0,0)$-M$(u,0,w)$-A$(u,0,\frac{1}{2})$}}&$2\mathrm{M}1(1),2\mathrm{M}2(1);\mathrm{M}1(1)\oplus\mathrm{M}2(1),\mathrm{M}1(1)\oplus\mathrm{M}2(1)$&Type a\\ \hline
54&{\bf\color{red}{G$(\frac{1}{2},0,w)$-M$(u,0,w)$-Z$(0,0,\frac{1}{2})$}}&$2\mathrm{M}1(1),2\mathrm{M}2(1);\mathrm{M}1(1)\oplus\mathrm{M}2(1),\mathrm{M}1(1)\oplus\mathrm{M}2(1)$&Type a\\ \hline
54&{\bf\color{red}{G$(\frac{1}{2},0,w)$-M$(u,0,w)$-A$(u,0,\frac{1}{2})$}}&$2\mathrm{M}1(1),2\mathrm{M}2(1);\mathrm{M}1(1)\oplus\mathrm{M}2(1),\mathrm{M}1(1)\oplus\mathrm{M}2(1)$&Type a\\ \hline
54&{\bf\color{red}{S$(\frac{1}{2},\frac{1}{2},0)$-N$(u,\frac{1}{2},w)$-T$(0,\frac{1}{2},\frac{1}{2})$}}&$2\mathrm{N}1(1),2\mathrm{N}2(1);\mathrm{N}1(1)\oplus\mathrm{N}2(1),\mathrm{N}1(1)\oplus\mathrm{N}2(1)$&Type a\\ \hline
54&{\bf\color{red}{S$(\frac{1}{2},\frac{1}{2},0)$-N$(u,\frac{1}{2},w)$-E$(u,\frac{1}{2},\frac{1}{2})$}}&$2\mathrm{N}1(1),2\mathrm{N}2(1);\mathrm{N}1(1)\oplus\mathrm{N}2(1),\mathrm{N}1(1)\oplus\mathrm{N}2(1)$&Type a\\
\hline
\hline \end{tabular}\end{table*}\end{center}

\newpage

\begin{center}\begin{table*}(Continued from previous table: Time-reversal symmetric, neglecting SOC)\\\begin{tabular}{c|c|c|c}
\hline \hline
SG& R-X-B & Splitting pattern in the  hourglss band structure & Type  \\
\hline

60&{\bf\color{red}{B$(0,v,\frac{1}{2})$-K$(0,v,w)$-Y$(0,\frac{1}{2},0)$}}&$2\mathrm{K}1(1),2\mathrm{K}2(1);\mathrm{K}1(1)\oplus\mathrm{K}2(1),\mathrm{K}1(1)\oplus\mathrm{K}2(1)$&Type a\\ \hline
60&{\bf\color{red}{B$(0,v,\frac{1}{2})$-K$(0,v,w)$-H$(0,\frac{1}{2},w)$}}&$2\mathrm{K}1(1),2\mathrm{K}2(1);\mathrm{K}1(1)\oplus\mathrm{K}2(1),\mathrm{K}1(1)\oplus\mathrm{K}2(1)$&Type a\\ \hline
60&{\bf\color{red}{X$(\frac{1}{2},0,0)$-M$(u,0,w)$-Z$(0,0,\frac{1}{2})$}}&$2\mathrm{M}1(1),2\mathrm{M}2(1);\mathrm{M}1(1)\oplus\mathrm{M}2(1),\mathrm{M}1(1)\oplus\mathrm{M}2(1)$&Type a\\ \hline
60&{\bf\color{red}{X$(\frac{1}{2},0,0)$-M$(u,0,w)$-A$(u,0,\frac{1}{2})$}}&$2\mathrm{M}1(1),2\mathrm{M}2(1);\mathrm{M}1(1)\oplus\mathrm{M}2(1),\mathrm{M}1(1)\oplus\mathrm{M}2(1)$&Type a\\ \hline
60&{\bf\color{red}{G$(\frac{1}{2},0,w)$-M$(u,0,w)$-Z$(0,0,\frac{1}{2})$}}&$2\mathrm{M}1(1),2\mathrm{M}2(1);\mathrm{M}1(1)\oplus\mathrm{M}2(1),\mathrm{M}1(1)\oplus\mathrm{M}2(1)$&Type a\\ \hline
60&{\bf\color{red}{G$(\frac{1}{2},0,w)$-M$(u,0,w)$-A$(u,0,\frac{1}{2})$}}&$2\mathrm{M}1(1),2\mathrm{M}2(1);\mathrm{M}1(1)\oplus\mathrm{M}2(1),\mathrm{M}1(1)\oplus\mathrm{M}2(1)$&Type a\\ \hline
60&{\bf\color{red}{E$(u,\frac{1}{2},\frac{1}{2})$-N$(u,\frac{1}{2},w)$-S$(\frac{1}{2},\frac{1}{2},0)$}}&$2\mathrm{N}1(1),2\mathrm{N}2(1);\mathrm{N}1(1)\oplus\mathrm{N}2(1),\mathrm{N}1(1)\oplus\mathrm{N}2(1)$&Type a\\
\hline
\hline \end{tabular}\end{table*}\end{center}

\newpage

\begin{center}\begin{table*}(Continued from previous table: Time-reversal symmetric, neglecting SOC)\\\begin{tabular}{c|c|c|c}
\hline \hline
SG& R-X-B &  Splitting pattern in the  hourglss band structure & Type  \\
\hline

86&\textbf{A$(\frac{1}{2},\frac{1}{2},\frac{1}{2})$-E$(u,v,\frac{1}{2})$-Z$(0,0,\frac{1}{2})$}&$2\mathrm{E}2(1),2\mathrm{E}1(1);\mathrm{E}1(1)\oplus\mathrm{E}2(1),\mathrm{E}1(1)\oplus\mathrm{E}2(1)$&Type a\\ \hline
86&\textbf{A$(\frac{1}{2},\frac{1}{2},\frac{1}{2})$-E$(u,v,\frac{1}{2})$-R$(0,\frac{1}{2},\frac{1}{2})$}&$2\mathrm{E}2(1),2\mathrm{E}1(1);\mathrm{E}1(1)\oplus\mathrm{E}2(1),\mathrm{E}1(1)\oplus\mathrm{E}2(1)$&Type a\\ \hline
86&\makecell{GM$(0,0,0)$-DT$(0,v,0)$-X$(0,\frac{1}{2},0)$\\D$(u,v,0)$}&$2\mathrm{DT}2(1),2\mathrm{DT}1(1);\mathrm{DT}1(1)\oplus\mathrm{DT}2(1),\mathrm{DT}1(1)\oplus\mathrm{DT}2(1)$&Type a\\ \hline
86&\makecell{A$(\frac{1}{2},\frac{1}{2},\frac{1}{2})$-S$(u,u,\frac{1}{2})$-Z$(0,0,\frac{1}{2})$\\E$(u,v,\frac{1}{2})$}&$2\mathrm{S}1(1),2\mathrm{S}2(1);\mathrm{S}1(1)\oplus\mathrm{S}2(1),\mathrm{S}1(1)\oplus\mathrm{S}2(1)$&Type a\\ \hline
86&\makecell{GM$(0,0,0)$-SM$(u,u,0)$-M$(\frac{1}{2},\frac{1}{2},0)$\\D$(u,v,0)$}&$2\mathrm{SM}2(1),2\mathrm{SM}1(1);\mathrm{SM}1(1)\oplus\mathrm{SM}2(1),\mathrm{SM}1(1)\oplus\mathrm{SM}2(1)$&Type a\\ \hline
86&\makecell{A$(\frac{1}{2},\frac{1}{2},\frac{1}{2})$-T$(u,\frac{1}{2},\frac{1}{2})$-R$(0,\frac{1}{2},\frac{1}{2})$\\E$(u,v,\frac{1}{2})$}&$2\mathrm{T}2(1),2\mathrm{T}1(1);\mathrm{T}1(1)\oplus\mathrm{T}2(1),\mathrm{T}1(1)\oplus\mathrm{T}2(1)$&Type a\\ \hline
88&\textbf{GM$(0,0,0)$-C$(u,v,0)$-M$(1,2,0)$}&$2\mathrm{C}2(1),2\mathrm{C}1(1);\mathrm{C}1(1)\oplus\mathrm{C}2(1),\mathrm{C}1(1)\oplus\mathrm{C}2(1)$&Type a\\
\hline
\hline \end{tabular}\end{table*}\end{center}

\newpage

\begin{center}\begin{table*}(Continued from previous table: Time-reversal symmetric, neglecting SOC)\\\begin{tabular}{c|c|c|c}
\hline \hline
SG& R-X-B &  Splitting pattern in the  hourglss band structure & Type  \\
\hline

101&\textbf{A$(\frac{1}{2},\frac{1}{2},\frac{1}{2})$-C$(u,u,w)$-M$(\frac{1}{2},\frac{1}{2},0)$}&$2\mathrm{C}1(1),2\mathrm{C}2(1);\mathrm{C}1(1)\oplus\mathrm{C}2(1),\mathrm{C}1(1)\oplus\mathrm{C}2(1)$&Type a\\ \hline
101&\textbf{A$(\frac{1}{2},\frac{1}{2},\frac{1}{2})$-C$(u,u,w)$-V$(\frac{1}{2},\frac{1}{2},w)$}&$2\mathrm{C}1(1),2\mathrm{C}2(1);\mathrm{C}1(1)\oplus\mathrm{C}2(1),\mathrm{C}1(1)\oplus\mathrm{C}2(1)$&Type a\\ \hline
101&\textbf{A$(\frac{1}{2},\frac{1}{2},\frac{1}{2})$-C$(u,u,w)$-Z$(0,0,\frac{1}{2})$}&$2\mathrm{C}1(1),2\mathrm{C}2(1);\mathrm{C}1(1)\oplus\mathrm{C}2(1),\mathrm{C}1(1)\oplus\mathrm{C}2(1)$&Type a\\ \hline
101&\textbf{Z$(0,0,\frac{1}{2})$-C$(u,u,w)$-GM$(0,0,0)$}&$2\mathrm{C}1(1),2\mathrm{C}2(1);\mathrm{C}1(1)\oplus\mathrm{C}2(1),\mathrm{C}1(1)\oplus\mathrm{C}2(1)$&Type a\\ \hline
101&\textbf{Z$(0,0,\frac{1}{2})$-C$(u,u,w)$-LD$(0,0,w)$}&$2\mathrm{C}1(1),2\mathrm{C}2(1);\mathrm{C}1(1)\oplus\mathrm{C}2(1),\mathrm{C}1(1)\oplus\mathrm{C}2(1)$&Type a\\ \hline
101&\textbf{Z$(0,0,\frac{1}{2})$-C$(u,u,w)$-M$(\frac{1}{2},\frac{1}{2},0)$}&$2\mathrm{C}1(1),2\mathrm{C}2(1);\mathrm{C}1(1)\oplus\mathrm{C}2(1),\mathrm{C}1(1)\oplus\mathrm{C}2(1)$&Type a\\ \hline
101&\textbf{Z$(0,0,\frac{1}{2})$-C$(u,u,w)$-V$(\frac{1}{2},\frac{1}{2},w)$}&$2\mathrm{C}1(1),2\mathrm{C}2(1);\mathrm{C}1(1)\oplus\mathrm{C}2(1),\mathrm{C}1(1)\oplus\mathrm{C}2(1)$&Type a\\
\hline
\hline \end{tabular}\end{table*}\end{center}

\newpage

\begin{center}\begin{table*}(Continued from previous table: Time-reversal symmetric, neglecting SOC)\\\begin{tabular}{c|c|c|c}
\hline \hline
SG& R-X-B &    Splitting pattern in the  hourglss band structure & Type  \\
\hline

104&\makecell{M$(\frac{1}{2},\frac{1}{2},0)$-SM$(u,u,0)$-GM$(0,0,0)$\\C$(u,u,w)$}&$2\mathrm{SM}1(1),2\mathrm{SM}2(1);\mathrm{SM}1(1)\oplus\mathrm{SM}2(1),\mathrm{SM}1(1)\oplus\mathrm{SM}2(1)$&Type a\\ \hline
104&\makecell{A$(\frac{1}{2},\frac{1}{2},\frac{1}{2})$-T$(u,\frac{1}{2},\frac{1}{2})$-R$(0,\frac{1}{2},\frac{1}{2})$\\F$(u,\frac{1}{2},w)$}&$2\mathrm{T}2(1),2\mathrm{T}1(1);\mathrm{T}1(1)\oplus\mathrm{T}2(1),\mathrm{T}1(1)\oplus\mathrm{T}2(1)$&Type a\\ \hline
104&\makecell{A$(\frac{1}{2},\frac{1}{2},\frac{1}{2})$-V$(\frac{1}{2},\frac{1}{2},w)$-M$(\frac{1}{2},\frac{1}{2},0)$\\C$(u,u,w)$$\mid$F$(u,\frac{1}{2},w)$}&$\mathrm{V}1(1)\oplus\mathrm{V}2(1),\mathrm{V}3(1)\oplus\mathrm{V}4(1);\mathrm{V}1(1)\oplus\mathrm{V}3(1),\mathrm{V}2(1)\oplus\mathrm{V}4(1)$&Type c\\ \hline
105&\textbf{Z$(0,0,\frac{1}{2})$-B$(0,v,w)$-GM$(0,0,0)$}&$2\mathrm{B}1(1),2\mathrm{B}2(1);\mathrm{B}1(1)\oplus\mathrm{B}2(1),\mathrm{B}1(1)\oplus\mathrm{B}2(1)$&Type a\\ \hline
105&\textbf{Z$(0,0,\frac{1}{2})$-B$(0,v,w)$-LD$(0,0,w)$}&$2\mathrm{B}1(1),2\mathrm{B}2(1);\mathrm{B}1(1)\oplus\mathrm{B}2(1),\mathrm{B}1(1)\oplus\mathrm{B}2(1)$&Type a\\ \hline
105&\textbf{A$(\frac{1}{2},\frac{1}{2},\frac{1}{2})$-F$(u,\frac{1}{2},w)$-M$(\frac{1}{2},\frac{1}{2},0)$}&$2\mathrm{F}1(1),2\mathrm{F}2(1);\mathrm{F}1(1)\oplus\mathrm{F}2(1),\mathrm{F}1(1)\oplus\mathrm{F}2(1)$&Type a\\ \hline
105&\textbf{A$(\frac{1}{2},\frac{1}{2},\frac{1}{2})$-F$(u,\frac{1}{2},w)$-V$(\frac{1}{2},\frac{1}{2},w)$}&$2\mathrm{F}1(1),2\mathrm{F}2(1);\mathrm{F}1(1)\oplus\mathrm{F}2(1),\mathrm{F}1(1)\oplus\mathrm{F}2(1)$&Type a\\
\hline
\hline \end{tabular}\end{table*}\end{center}

\newpage

\begin{center}\begin{table*}(Continued from previous table: Time-reversal symmetric, neglecting SOC)\\\begin{tabular}{c|c|c|c}
\hline \hline
SG& R-X-B &  Splitting pattern in the  hourglss band structure & Type  \\
\hline

113&\textbf{M$(\frac{1}{2},\frac{1}{2},0)$-C$(u,u,w)$-V$(\frac{1}{2},\frac{1}{2},w)$}&$\mathrm{C}1(1)\oplus\mathrm{C}2(1),2\mathrm{C}1(1);2\mathrm{C}1(1),\mathrm{C}1(1)\oplus\mathrm{C}2(1)$&Type d\\ \hline
113&\textbf{A$(\frac{1}{2},\frac{1}{2},\frac{1}{2})$-C$(u,u,w)$-GM$(0,0,0)$}&$2\mathrm{C}1(1),2\mathrm{C}2(1);\mathrm{C}1(1)\oplus\mathrm{C}2(1),\mathrm{C}1(1)\oplus\mathrm{C}2(1)$&Type a\\ \hline
113&\textbf{A$(\frac{1}{2},\frac{1}{2},\frac{1}{2})$-C$(u,u,w)$-M$(\frac{1}{2},\frac{1}{2},0)$}&$2\mathrm{C}1(1),2\mathrm{C}2(1);\mathrm{C}1(1)\oplus\mathrm{C}2(1),\mathrm{C}1(1)\oplus\mathrm{C}2(1)$&Type a\\ \hline
113&\textbf{A$(\frac{1}{2},\frac{1}{2},\frac{1}{2})$-C$(u,u,w)$-Z$(0,0,\frac{1}{2})$}&$2\mathrm{C}1(1),2\mathrm{C}2(1);\mathrm{C}1(1)\oplus\mathrm{C}2(1),\mathrm{C}1(1)\oplus\mathrm{C}2(1)$&Type a\\ \hline
113&\textbf{A$(\frac{1}{2},\frac{1}{2},\frac{1}{2})$-C$(u,u,w)$-LD$(0,0,w)$}&$2\mathrm{C}1(1),2\mathrm{C}2(1);\mathrm{C}1(1)\oplus\mathrm{C}2(1),\mathrm{C}1(1)\oplus\mathrm{C}2(1)$&Type a\\ \hline
113&\textbf{A$(\frac{1}{2},\frac{1}{2},\frac{1}{2})$-C$(u,u,w)$-V$(\frac{1}{2},\frac{1}{2},w)$}&$2\mathrm{C}1(1),2\mathrm{C}2(1);\mathrm{C}1(1)\oplus\mathrm{C}2(1),\mathrm{C}1(1)\oplus\mathrm{C}2(1)$&Type a\\ \hline
113&\textbf{M$(\frac{1}{2},\frac{1}{2},0)$-C$(u,u,w)$-GM$(0,0,0)$}&$2\mathrm{C}1(1),2\mathrm{C}2(1);\mathrm{C}1(1)\oplus\mathrm{C}2(1),\mathrm{C}1(1)\oplus\mathrm{C}2(1)$&Type a\\
\hline
\hline \end{tabular}\end{table*}\end{center}

\newpage

\begin{center}\begin{table*}(Continued from previous table: Time-reversal symmetric, neglecting SOC)\\\begin{tabular}{c|c|c|c}
\hline \hline
SG& R-X-B &  Splitting pattern in the  hourglss band structure & Type  \\
\hline

125&\textbf{A$(\frac{1}{2},\frac{1}{2},\frac{1}{2})$-C$(u,u,w)$-M$(\frac{1}{2},\frac{1}{2},0)$}&$\mathrm{C}1(1)\oplus\mathrm{C}2(1),2\mathrm{C}1(1);2\mathrm{C}1(1),\mathrm{C}1(1)\oplus\mathrm{C}2(1)$&Type d\\ \hline
125&\textbf{A$(\frac{1}{2},\frac{1}{2},\frac{1}{2})$-C$(u,u,w)$-M$(\frac{1}{2},\frac{1}{2},0)$}&$\mathrm{C}1(1)\oplus\mathrm{C}2(1),2\mathrm{C}2(1);2\mathrm{C}2(1),\mathrm{C}1(1)\oplus\mathrm{C}2(1)$&Type d\\ \hline
125&\textbf{A$(\frac{1}{2},\frac{1}{2},\frac{1}{2})$-C$(u,u,w)$-GM$(0,0,0)$}&$2\mathrm{C}1(1),2\mathrm{C}2(1);\mathrm{C}1(1)\oplus\mathrm{C}2(1),\mathrm{C}1(1)\oplus\mathrm{C}2(1)$&Type a\\ \hline
125&\textbf{A$(\frac{1}{2},\frac{1}{2},\frac{1}{2})$-C$(u,u,w)$-M$(\frac{1}{2},\frac{1}{2},0)$}&$2\mathrm{C}1(1),2\mathrm{C}2(1);\mathrm{C}1(1)\oplus\mathrm{C}2(1),\mathrm{C}1(1)\oplus\mathrm{C}2(1)$&Type a\\ \hline
125&\textbf{A$(\frac{1}{2},\frac{1}{2},\frac{1}{2})$-C$(u,u,w)$-Z$(0,0,\frac{1}{2})$}&$2\mathrm{C}1(1),2\mathrm{C}2(1);\mathrm{C}1(1)\oplus\mathrm{C}2(1),\mathrm{C}1(1)\oplus\mathrm{C}2(1)$&Type a\\ \hline
125&\textbf{A$(\frac{1}{2},\frac{1}{2},\frac{1}{2})$-C$(u,u,w)$-LD$(0,0,w)$}&$2\mathrm{C}1(1),2\mathrm{C}2(1);\mathrm{C}1(1)\oplus\mathrm{C}2(1),\mathrm{C}1(1)\oplus\mathrm{C}2(1)$&Type a\\ \hline
125&\textbf{A$(\frac{1}{2},\frac{1}{2},\frac{1}{2})$-C$(u,u,w)$-V$(\frac{1}{2},\frac{1}{2},w)$}&$2\mathrm{C}1(1),2\mathrm{C}2(1);\mathrm{C}1(1)\oplus\mathrm{C}2(1),\mathrm{C}1(1)\oplus\mathrm{C}2(1)$&Type a\\
\hline
\hline \end{tabular}\end{table*}\end{center}

\newpage

\begin{center}\begin{table*}(Continued from previous table: Time-reversal symmetric, neglecting SOC)\\\begin{tabular}{c|c|c|c}
\hline \hline
SG& R-X-B &  Splitting pattern in the  hourglss band structure & Type  \\
\hline

126&\textbf{A$(\frac{1}{2},\frac{1}{2},\frac{1}{2})$-F$(u,\frac{1}{2},w)$-X$(0,\frac{1}{2},0)$}&$2\mathrm{F}2(1),2\mathrm{F}1(1);\mathrm{F}1(1)\oplus\mathrm{F}2(1),\mathrm{F}1(1)\oplus\mathrm{F}2(1)$&Type a\\ \hline
126&\textbf{A$(\frac{1}{2},\frac{1}{2},\frac{1}{2})$-F$(u,\frac{1}{2},w)$-W$(0,\frac{1}{2},w)$}&$2\mathrm{F}2(1),2\mathrm{F}1(1);\mathrm{F}1(1)\oplus\mathrm{F}2(1),\mathrm{F}1(1)\oplus\mathrm{F}2(1)$&Type a\\ \hline
126&\textbf{A$(\frac{1}{2},\frac{1}{2},\frac{1}{2})$-F$(u,\frac{1}{2},w)$-Y$(u,\frac{1}{2},0)$}&$2\mathrm{F}2(1),2\mathrm{F}1(1);\mathrm{F}1(1)\oplus\mathrm{F}2(1),\mathrm{F}1(1)\oplus\mathrm{F}2(1)$&Type a\\ \hline
126&\makecell{A$(\frac{1}{2},\frac{1}{2},\frac{1}{2})$-V$(\frac{1}{2},\frac{1}{2},w)$-M$(\frac{1}{2},\frac{1}{2},0)$\\C$(u,u,w)$$\mid$F$(u,\frac{1}{2},w)$}&$\mathrm{V}1(1)\oplus\mathrm{V}2(1),\mathrm{V}3(1)\oplus\mathrm{V}4(1);\mathrm{V}1(1)\oplus\mathrm{V}3(1),\mathrm{V}2(1)\oplus\mathrm{V}4(1)$&Type c\\ \hline
126&\makecell{GM$(0,0,0)$-DT$(0,v,0)$-X$(0,\frac{1}{2},0)$\\D$(u,v,0)$}&$\mathrm{DT}2(1)\oplus\mathrm{DT}3(1),\mathrm{DT}1(1)\oplus\mathrm{DT}4(1);\mathrm{DT}1(1)\oplus\mathrm{DT}2(1),\mathrm{DT}3(1)\oplus\mathrm{DT}4(1)$&Type c\\ \hline
126&\makecell{GM$(0,0,0)$-SM$(u,u,0)$-M$(\frac{1}{2},\frac{1}{2},0)$\\D$(u,v,0)$}&$\mathrm{SM}3(1)\oplus\mathrm{SM}4(1),\mathrm{SM}1(1)\oplus\mathrm{SM}2(1);\mathrm{SM}1(1)\oplus\mathrm{SM}4(1),\mathrm{SM}2(1)\oplus\mathrm{SM}3(1)$&Type c\\ \hline
126&\makecell{GM$(0,0,0)$-SM$(u,u,0)$-M$(\frac{1}{2},\frac{1}{2},0)$\\C$(u,u,w)$$\mid$D$(u,v,0)$}&$\mathrm{SM}3(1)\oplus\mathrm{SM}4(1),\mathrm{SM}1(1)\oplus\mathrm{SM}2(1);\mathrm{SM}1(1)\oplus\mathrm{SM}3(1),\mathrm{SM}2(1)\oplus\mathrm{SM}4(1)$&Type c\\
\hline
\hline \end{tabular}\end{table*}\end{center}

\newpage

\begin{center}\begin{table*}(Continued from previous table: Time-reversal symmetric, neglecting SOC)\\\begin{tabular}{c|c|c|c}
\hline \hline
SG& R-X-B &  Splitting pattern in the  hourglss band structure & Type  \\
\hline

128&\textbf{M$(\frac{1}{2},\frac{1}{2},0)$-C$(u,u,w)$-Z$(0,0,\frac{1}{2})$}&$2\mathrm{C}1(1),2\mathrm{C}2(1);\mathrm{C}1(1)\oplus\mathrm{C}2(1),\mathrm{C}1(1)\oplus\mathrm{C}2(1)$&Type a\\ \hline
128&\textbf{M$(\frac{1}{2},\frac{1}{2},0)$-C$(u,u,w)$-LD$(0,0,w)$}&$2\mathrm{C}1(1),2\mathrm{C}2(1);\mathrm{C}1(1)\oplus\mathrm{C}2(1),\mathrm{C}1(1)\oplus\mathrm{C}2(1)$&Type a\\ \hline
128&\textbf{M$(\frac{1}{2},\frac{1}{2},0)$-C$(u,u,w)$-V$(\frac{1}{2},\frac{1}{2},w)$}&$2\mathrm{C}1(1),2\mathrm{C}2(1);\mathrm{C}1(1)\oplus\mathrm{C}2(1),\mathrm{C}1(1)\oplus\mathrm{C}2(1)$&Type a\\ \hline
128&\textbf{M$(\frac{1}{2},\frac{1}{2},0)$-C$(u,u,w)$-S$(u,u,\frac{1}{2})$}&$2\mathrm{C}1(1),2\mathrm{C}2(1);\mathrm{C}1(1)\oplus\mathrm{C}2(1),\mathrm{C}1(1)\oplus\mathrm{C}2(1)$&Type a\\ \hline
128&\textbf{V$(\frac{1}{2},\frac{1}{2},w)$-C$(u,u,w)$-A$(\frac{1}{2},\frac{1}{2},\frac{1}{2})$}&$2\mathrm{C}1(1),2\mathrm{C}2(1);\mathrm{C}1(1)\oplus\mathrm{C}2(1),\mathrm{C}1(1)\oplus\mathrm{C}2(1)$&Type a\\ \hline
128&\textbf{V$(\frac{1}{2},\frac{1}{2},w)$-C$(u,u,w)$-GM$(0,0,0)$}&$2\mathrm{C}1(1),2\mathrm{C}2(1);\mathrm{C}1(1)\oplus\mathrm{C}2(1),\mathrm{C}1(1)\oplus\mathrm{C}2(1)$&Type a\\ \hline
128&\textbf{V$(\frac{1}{2},\frac{1}{2},w)$-C$(u,u,w)$-Z$(0,0,\frac{1}{2})$}&$2\mathrm{C}1(1),2\mathrm{C}2(1);\mathrm{C}1(1)\oplus\mathrm{C}2(1),\mathrm{C}1(1)\oplus\mathrm{C}2(1)$&Type a\\
\hline
\hline \end{tabular}\end{table*}\end{center}

\newpage

\begin{center}\begin{table*}(Continued from previous table: Time-reversal symmetric, neglecting SOC)\\\begin{tabular}{c|c|c|c}
\hline \hline
SG& R-X-B &  Splitting pattern in the  hourglss band structure & Type  \\
\hline

129&\textbf{Z$(0,0,\frac{1}{2})$-E$(u,v,\frac{1}{2})$-A$(\frac{1}{2},\frac{1}{2},\frac{1}{2})$}&$2\mathrm{E}2(1),2\mathrm{E}1(1);\mathrm{E}1(1)\oplus\mathrm{E}2(1),\mathrm{E}1(1)\oplus\mathrm{E}2(1)$&Type a\\ \hline
129&\textbf{Z$(0,0,\frac{1}{2})$-E$(u,v,\frac{1}{2})$-R$(0,\frac{1}{2},\frac{1}{2})$}&$2\mathrm{E}2(1),2\mathrm{E}1(1);\mathrm{E}1(1)\oplus\mathrm{E}2(1),\mathrm{E}1(1)\oplus\mathrm{E}2(1)$&Type a\\ \hline
129&\textbf{Z$(0,0,\frac{1}{2})$-E$(u,v,\frac{1}{2})$-T$(u,\frac{1}{2},\frac{1}{2})$}&$2\mathrm{E}2(1),2\mathrm{E}1(1);\mathrm{E}1(1)\oplus\mathrm{E}2(1),\mathrm{E}1(1)\oplus\mathrm{E}2(1)$&Type a\\ \hline
129&\makecell{GM$(0,0,0)$-DT$(0,v,0)$-X$(0,\frac{1}{2},0)$\\B$(0,v,w)$$\mid$D$(u,v,0)$}&$\mathrm{DT}2(1)\oplus\mathrm{DT}3(1),\mathrm{DT}1(1)\oplus\mathrm{DT}4(1);\mathrm{DT}1(1)\oplus\mathrm{DT}3(1),\mathrm{DT}2(1)\oplus\mathrm{DT}4(1)$&Type c\\ \hline
129&\makecell{A$(\frac{1}{2},\frac{1}{2},\frac{1}{2})$-S$(u,u,\frac{1}{2})$-Z$(0,0,\frac{1}{2})$\\E$(u,v,\frac{1}{2})$}&$\mathrm{S}2(1)\oplus\mathrm{S}3(1),\mathrm{S}1(1)\oplus\mathrm{S}4(1);\mathrm{S}1(1)\oplus\mathrm{S}2(1),\mathrm{S}3(1)\oplus\mathrm{S}4(1)$&Type c\\ \hline
129&\makecell{A$(\frac{1}{2},\frac{1}{2},\frac{1}{2})$-S$(u,u,\frac{1}{2})$-Z$(0,0,\frac{1}{2})$\\C$(u,u,w)$$\mid$E$(u,v,\frac{1}{2})$}&$\mathrm{S}1(1)\oplus\mathrm{S}3(1),\mathrm{S}2(1)\oplus\mathrm{S}4(1);\mathrm{S}1(1)\oplus\mathrm{S}2(1),\mathrm{S}3(1)\oplus\mathrm{S}4(1)$&Type c\\ \hline
129&\makecell{GM$(0,0,0)$-SM$(u,u,0)$-M$(\frac{1}{2},\frac{1}{2},0)$\\D$(u,v,0)$}&$\mathrm{SM}3(1)\oplus\mathrm{SM}4(1),\mathrm{SM}1(1)\oplus\mathrm{SM}2(1);\mathrm{SM}1(1)\oplus\mathrm{SM}4(1),\mathrm{SM}2(1)\oplus\mathrm{SM}3(1)$&Type c\\
\hline
\hline \end{tabular}\end{table*}\end{center}

\newpage

\begin{center}\begin{table*}(Continued from previous table: Time-reversal symmetric, neglecting SOC)\\\begin{tabular}{c|c|c|c}
\hline \hline
SG& R-X-B &    Splitting pattern in the  hourglss band structure & Type  \\
\hline

130&\makecell{GM$(0,0,0)$-DT$(0,v,0)$-X$(0,\frac{1}{2},0)$\\B$(0,v,w)$$\mid$D$(u,v,0)$}&$\mathrm{DT}2(1)\oplus\mathrm{DT}3(1),\mathrm{DT}1(1)\oplus\mathrm{DT}4(1);\mathrm{DT}1(1)\oplus\mathrm{DT}3(1),\mathrm{DT}2(1)\oplus\mathrm{DT}4(1)$&Type c\\ \hline
130&\makecell{GM$(0,0,0)$-SM$(u,u,0)$-M$(\frac{1}{2},\frac{1}{2},0)$\\D$(u,v,0)$}&$\mathrm{SM}3(1)\oplus\mathrm{SM}4(1),\mathrm{SM}1(1)\oplus\mathrm{SM}2(1);\mathrm{SM}1(1)\oplus\mathrm{SM}4(1),\mathrm{SM}2(1)\oplus\mathrm{SM}3(1)$&Type c\\ \hline
130&\makecell{GM$(0,0,0)$-SM$(u,u,0)$-M$(\frac{1}{2},\frac{1}{2},0)$\\C$(u,u,w)$$\mid$D$(u,v,0)$}&$\mathrm{SM}3(1)\oplus\mathrm{SM}4(1),\mathrm{SM}1(1)\oplus\mathrm{SM}2(1);\mathrm{SM}1(1)\oplus\mathrm{SM}3(1),\mathrm{SM}2(1)\oplus\mathrm{SM}4(1)$&Type c\\ \hline
131&\textbf{Z$(0,0,\frac{1}{2})$-B$(0,v,w)$-GM$(0,0,0)$}&$2\mathrm{B}1(1),2\mathrm{B}2(1);\mathrm{B}1(1)\oplus\mathrm{B}2(1),\mathrm{B}1(1)\oplus\mathrm{B}2(1)$&Type a\\ \hline
131&\textbf{Z$(0,0,\frac{1}{2})$-B$(0,v,w)$-LD$(0,0,w)$}&$2\mathrm{B}1(1),2\mathrm{B}2(1);\mathrm{B}1(1)\oplus\mathrm{B}2(1),\mathrm{B}1(1)\oplus\mathrm{B}2(1)$&Type a\\ \hline
131&\textbf{A$(\frac{1}{2},\frac{1}{2},\frac{1}{2})$-F$(u,\frac{1}{2},w)$-M$(\frac{1}{2},\frac{1}{2},0)$}&$2\mathrm{F}1(1),2\mathrm{F}2(1);\mathrm{F}1(1)\oplus\mathrm{F}2(1),\mathrm{F}1(1)\oplus\mathrm{F}2(1)$&Type a\\ \hline
131&\textbf{A$(\frac{1}{2},\frac{1}{2},\frac{1}{2})$-F$(u,\frac{1}{2},w)$-V$(\frac{1}{2},\frac{1}{2},w)$}&$2\mathrm{F}1(1),2\mathrm{F}2(1);\mathrm{F}1(1)\oplus\mathrm{F}2(1),\mathrm{F}1(1)\oplus\mathrm{F}2(1)$&Type a\\
\hline
\hline \end{tabular}\end{table*}\end{center}

\newpage

\begin{center}\begin{table*}(Continued from previous table: Time-reversal symmetric, neglecting SOC)\\\begin{tabular}{c|c|c|c}
\hline \hline
SG& R-X-B &  Splitting pattern in the  hourglss band structure & Type  \\
\hline

133&\textbf{M$(\frac{1}{2},\frac{1}{2},0)$-C$(u,u,w)$-V$(\frac{1}{2},\frac{1}{2},w)$}&$2\mathrm{C}1(1),2\mathrm{C}2(1);\mathrm{C}1(1)\oplus\mathrm{C}2(1),\mathrm{C}1(1)\oplus\mathrm{C}2(1)$&Type a\\ \hline
133&\textbf{M$(\frac{1}{2},\frac{1}{2},0)$-C$(u,u,w)$-S$(u,u,\frac{1}{2})$}&$2\mathrm{C}1(1),2\mathrm{C}2(1);\mathrm{C}1(1)\oplus\mathrm{C}2(1),\mathrm{C}1(1)\oplus\mathrm{C}2(1)$&Type a\\ \hline
133&\textbf{GM$(0,0,0)$-D$(u,v,0)$-M$(\frac{1}{2},\frac{1}{2},0)$}&$2\mathrm{D}2(1),2\mathrm{D}1(1);\mathrm{D}1(1)\oplus\mathrm{D}2(1),\mathrm{D}1(1)\oplus\mathrm{D}2(1)$&Type a\\ \hline
133&\textbf{GM$(0,0,0)$-D$(u,v,0)$-X$(0,\frac{1}{2},0)$}&$2\mathrm{D}2(1),2\mathrm{D}1(1);\mathrm{D}1(1)\oplus\mathrm{D}2(1),\mathrm{D}1(1)\oplus\mathrm{D}2(1)$&Type a\\ \hline
133&\textbf{GM$(0,0,0)$-D$(u,v,0)$-Y$(u,\frac{1}{2},0)$}&$2\mathrm{D}2(1),2\mathrm{D}1(1);\mathrm{D}1(1)\oplus\mathrm{D}2(1),\mathrm{D}1(1)\oplus\mathrm{D}2(1)$&Type a\\ \hline
133&\makecell{A$(\frac{1}{2},\frac{1}{2},\frac{1}{2})$-V$(\frac{1}{2},\frac{1}{2},w)$-M$(\frac{1}{2},\frac{1}{2},0)$\\C$(u,u,w)$}&$\mathrm{V}2(1)\oplus\mathrm{V}3(1),\mathrm{V}1(1)\oplus\mathrm{V}4(1);\mathrm{V}1(1)\oplus\mathrm{V}3(1),\mathrm{V}2(1)\oplus\mathrm{V}4(1)$&Type c\\ \hline
133&\makecell{GM$(0,0,0)$-DT$(0,v,0)$-X$(0,\frac{1}{2},0)$\\D$(u,v,0)$}&$\mathrm{DT}2(1)\oplus\mathrm{DT}3(1),\mathrm{DT}1(1)\oplus\mathrm{DT}4(1);\mathrm{DT}1(1)\oplus\mathrm{DT}2(1),\mathrm{DT}3(1)\oplus\mathrm{DT}4(1)$&Type c\\
\hline
\hline \end{tabular}\end{table*}\end{center}

\newpage

\begin{center}\begin{table*}(Continued from previous table: Time-reversal symmetric, neglecting SOC)\\\begin{tabular}{c|c|c|c}
\hline \hline
SG& R-X-B &    Splitting pattern in the  hourglss band structure & Type  \\
\hline

134&\makecell{GM$(0,0,0)$-SM$(u,u,0)$-M$(\frac{1}{2},\frac{1}{2},0)$\\C$(u,u,w)$$\mid$D$(u,v,0)$}&$\mathrm{SM}3(1)\oplus\mathrm{SM}4(1),\mathrm{SM}1(1)\oplus\mathrm{SM}2(1);\mathrm{SM}1(1)\oplus\mathrm{SM}3(1),\mathrm{SM}2(1)\oplus\mathrm{SM}4(1)$&Type c\\ \hline
134&\makecell{A$(\frac{1}{2},\frac{1}{2},\frac{1}{2})$-T$(u,\frac{1}{2},\frac{1}{2})$-R$(0,\frac{1}{2},\frac{1}{2})$\\E$(u,v,\frac{1}{2})$}&$\mathrm{T}2(1)\oplus\mathrm{T}3(1),\mathrm{T}1(1)\oplus\mathrm{T}4(1);\mathrm{T}1(1)\oplus\mathrm{T}2(1),\mathrm{T}3(1)\oplus\mathrm{T}4(1)$&Type c\\ \hline
135&{\bf\color{red}{R$(0,\frac{1}{2},\frac{1}{2})$-E$(u,v,\frac{1}{2})$-Z$(0,0,\frac{1}{2})$}}&$2\mathrm{E}2(1),2\mathrm{E}1(1);\mathrm{E}1(1)\oplus\mathrm{E}2(1),\mathrm{E}1(1)\oplus\mathrm{E}2(1)$&Type a\\ \hline
135&{\bf\color{red}{R$(0,\frac{1}{2},\frac{1}{2})$-E$(u,v,\frac{1}{2})$-S$(u,u,\frac{1}{2})$}}&$2\mathrm{E}2(1),2\mathrm{E}1(1);\mathrm{E}1(1)\oplus\mathrm{E}2(1),\mathrm{E}1(1)\oplus\mathrm{E}2(1)$&Type a\\ \hline
135&{\bf\color{red}{T$(u,\frac{1}{2},\frac{1}{2})$-E$(u,v,\frac{1}{2})$-Z$(0,0,\frac{1}{2})$}}&$2\mathrm{E}1(1),2\mathrm{E}2(1);\mathrm{E}1(1)\oplus\mathrm{E}2(1),\mathrm{E}1(1)\oplus\mathrm{E}2(1)$&Type a\\ \hline
135&{\bf\color{red}{T$(u,\frac{1}{2},\frac{1}{2})$-E$(u,v,\frac{1}{2})$-S$(u,u,\frac{1}{2})$}}&$2\mathrm{E}1(1),2\mathrm{E}2(1);\mathrm{E}1(1)\oplus\mathrm{E}2(1),\mathrm{E}1(1)\oplus\mathrm{E}2(1)$&Type a\\ \hline
135&\textbf{M$(\frac{1}{2},\frac{1}{2},0)$-C$(u,u,w)$-V$(\frac{1}{2},\frac{1}{2},w)$}&$\mathrm{C}1(1)\oplus\mathrm{C}2(1),2\mathrm{C}1(1);2\mathrm{C}1(1),\mathrm{C}1(1)\oplus\mathrm{C}2(1)$&Type d\\
\hline
\hline \end{tabular}%
\end{table*}\end{center}

\newpage

\section{Hourglass band crossings for materials without  TRS and significant SOC}\label{notrssoc}
In this section, we list all possible positions in the BZ for 230 space groups  for systems without TRS but with significant SOC.
\begin{center}\begin{table*}%
\end{table*}\end{center}

\newpage

\begin{center}\begin{table*}(Continued from previous table: Time-reversal symmetry broken, considering SOC)\\%
\end{table*}\end{center}

\newpage

\begin{center}\begin{table*}(Continued from previous table: Time-reversal symmetry broken, considering SOC)\\%
\end{table*}\end{center}

\newpage

\begin{center}\begin{table*}(Continued from previous table: Time-reversal symmetry broken, considering SOC)\\%
\end{table*}\end{center}

\newpage
\section{Hourglass band crossings for materials without  TRS and negligible SOC}\label{notrsnosoc}
In this section, we list all possible positions in the BZ for 230 space groups  for systems in the absence of both TRS and SOC.
\begin{center}\begin{table*}%
\end{table*}\end{center}

\newpage

\begin{center}\begin{table*}(Continued from previous table: Time-reversal symmetry broken, neglecting SOC)\\%
\end{table*}\end{center}

\newpage

\begin{center}\begin{table*}(Continued from previous table: Time-reversal symmetry broken, neglecting SOC)\\%
\end{table*}\end{center}

\newpage

\begin{center}\begin{table*}(Continued from previous table: Time-reversal symmetry broken, neglecting SOC)\\%
\end{table*}\end{center}

\newpage
\section{The Other Material Representatives}\label{material}
\begin{figure*}[tbh!!]
	\includegraphics[width=0.5 \textwidth]{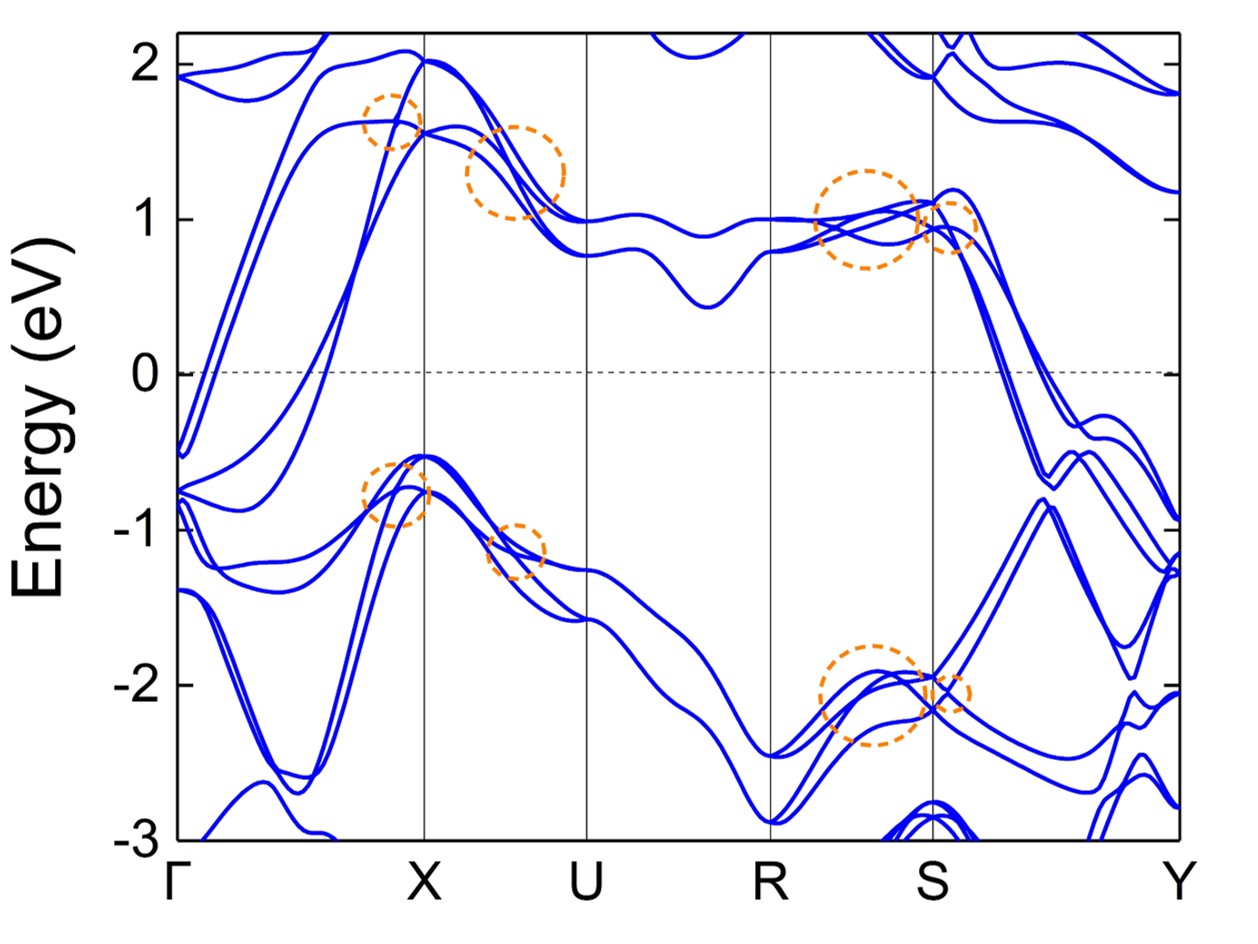}\\
	\caption{\textbf{First principles calculated band structure of OsN in SG 31 in the presence of SOC. The type $a$ hourglass crossing points appear along the X-SM-GM and S-C-Y line, and type $c$ hourglass crossing points appear along the U-G-X and R-Q-S line. These hourglass nodal points are indicated by orange dashed circles.
	}}\label{31}
\end{figure*}

\begin{figure*}[tbh!!]
	\includegraphics[width=0.5 \textwidth]{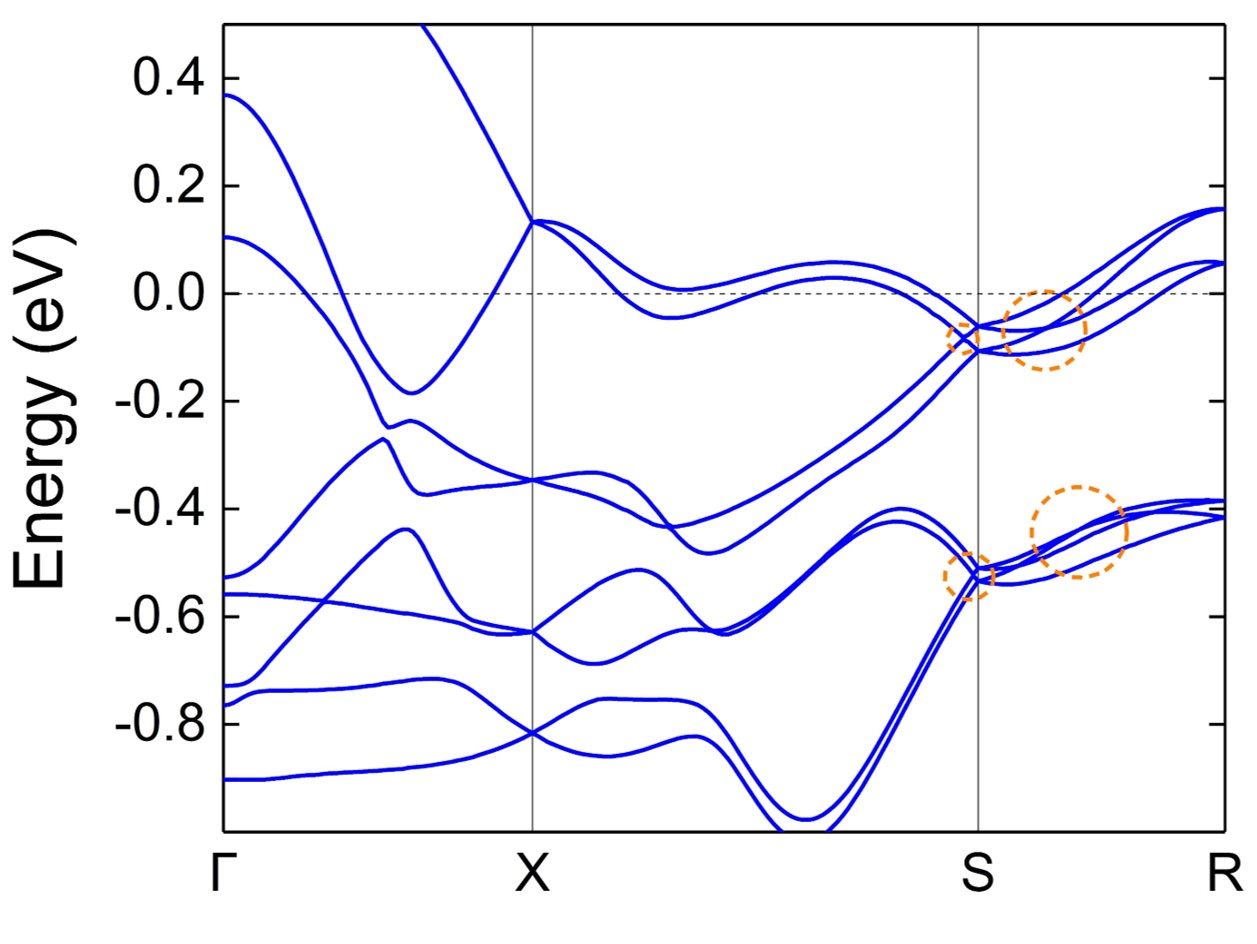}\\
	\caption{\textbf{First principles calculated band structure oof Y$_2$Au in SG 62 in the presence of SOC, which exhibit type $a$ hourglass crossing along the S-D-X line and type $c$ hourglass crossing along the R-Q-S line. The hourglass band crossings are denoted  by orange dashed circles.
	}}\label{y2au}
\end{figure*}

In this section, we present  two more hourglass materials within the space groups which have essential hourglass structures. Here we consider the situation with time-reversal symmetry and significant SOC. The positions of essential hourglass crossings indicated by red color in Sec. \ref{trssoc}. Here we present two materials to show the hourglass band structure. In Fig. \ref{31} we present the first-principles calculated electronic band structure of OsN \cite{osn}, which crystallizes in SG 31. We observe that there exist type $a$ hourglass band crossings in X($\frac{1}{2}$, 0, 0)-SM(u, 0, 0)-GM(0, 0, 0) and S($\frac{1}{2}$, $\frac{1}{2}$, 0)-C(u, $\frac{1}{2}$, 0)-Y(0, $\frac{1}{2}$, 0) line, and type $c$ hourglass crossing points appear along the U($\frac{1}{2}$, 0, $\frac{1}{2}$)-G($\frac{1}{2}$, 0, w)-X($\frac{1}{2}$, 0, 0) and  R($\frac{1}{2}$, $\frac{1}{2}$, $\frac{1}{2}$)-Q($\frac{1}{2}$, $\frac{1}{2}$, w)-S($\frac{1}{2}$, $\frac{1}{2}$, 0) line. The hourglass band crossing points are all double-fold degenerated and these band crossings can form two closed hourglass nodal loop enclosed X and S points within the $k_y$ =0 and $k_x$ =$\pi$ plane. In Fig. \ref{y2au} we displays the band structure near the Fermi level of the other hourglass material, Y$_2$Au\cite{y2au} by first-principles calculations. Y$_2$Au crystallizes in SG 62 and has the same symmetry as Ba$_2$ReO$_5$ shown in the main text. The same as Ba$_2$ReO$_5$,  Y$_2$Au also has one hourglass crossing of type $a$ along the S($\frac{1}{2}$, $\frac{1}{2}$, 0)-D($\frac{1}{2}$, v, 0)-X($\frac{1}{2}$, 0, 0) line, one  hourglass crossing of type $c$ along the R($\frac{1}{2}$, $\frac{1}{2}$, $\frac{1}{2}$)-Q($\frac{1}{2}$, $\frac{1}{2}$, w)-S($\frac{1}{2}$, $\frac{1}{2}$, 0) line and a closed nodal loop enclosed S point within the $k_x$ = $\pi$ plane.

\begin{figure*}[tbh]
	\includegraphics[width=1 \textwidth]{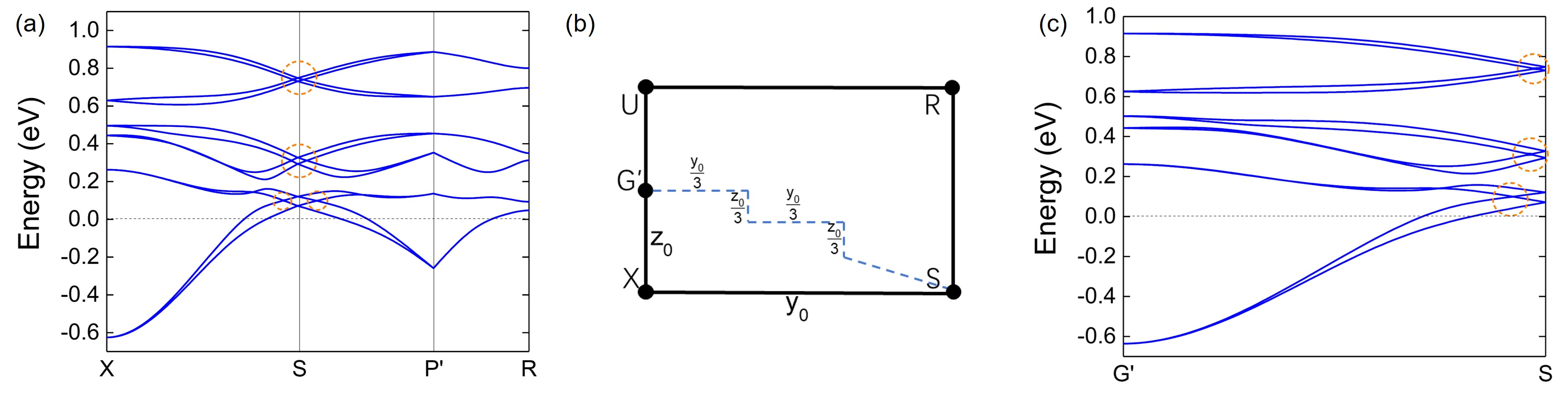}\\
	\caption{\textbf{(a) The electronic band structure of Ba$_2$ReO$_5$ where P$'$ is the middle point of U-R. Note that the bands in P$'$-R are always four-fold degenerate since only one 4 dimensional irrep is allowed in HSL P. (b)The schematic diagram of  an arbitrary path connecting S and G$'$, where G$'$ is the middle point of U-X: y$_0$ and z$_0$ are the length of X-S and X-G$'$. (c) The electronic band structure of G$'$-S along the path given in (b).
	}}\label{ba2reo5}
\end{figure*}

 With respect to Ba$_2$ReO$_5$,  we display the calculated  band structure along S-P$'$ and P$'$-R where P$'$ is the middle point of UR in Fig. \ref{ba2reo5}(a) .  It is observed that U-P-R are four-fold degenerate and no irreps splitting as the U-G-X line, and the line connecting S and any point in HSL  P could host essential hourglass band crossing.   Furthermore, we take an arbitrary path connecting G$'$ and S points, as shown in Fig. \ref{ba2reo5}(b) where y$_0$ nd z$_0$ are the length of X-S and X-G$'$. The electronic band structure of along that path  is shown in Fig. \ref{ba2reo5}(c) which hourglass band structures form as expected indicated by orange  dashed circles.

\section{Methods}
Our ab initio calculations are based on WIEN2K \cite{wien2k} using the linearized augmented plane wave (LAPW) method. We consider spin-orbital coupling in all of our calculations. The standard GGA with Perdew¨CBurke¨CErnzerhof (PBE) realization was adopted for the exchange-correlation functional \cite{gga}.
\clearpage
\end{document}